\documentclass[a4paper,11pt]{article}
\pdfoutput=1 

\usepackage{jheppub} 

\usepackage[T1]{fontenc} 
\usepackage[english]{babel}
\usepackage{graphicx}			    	
\usepackage{slashed}
\usepackage{caption}
\usepackage{amsmath}
\usepackage{amsthm}
\usepackage{amsfonts}
\usepackage{mathtools}
\usepackage{subcaption}
\usepackage{tikz} 
\usepackage{tikz-feynman}
\usepackage{tikz-3dplot}
\usepackage{float}
\tikzfeynmanset{compat=1.1.0}
\usepackage{placeins}
\usepackage{scalerel}
\usepackage{verbatim}
\usepackage{url}
\usetikzlibrary{shapes.misc}
\tikzset{cross/.style={cross out, draw=black, minimum size=2*(#1-\pgflinewidth), inner sep=0pt, outer sep=0pt},
	cross/.default={5pt}}

\DeclareMathOperator{\Det}{Det}

\DeclareMathOperator{\Tr}{Tr}

\newcommand*\Laplace{\mathop{}\!\mathbin\bigtriangleup}

\theoremstyle{plain}
\theoremstyle{plain}

\theoremstyle{definition}

\makeatletter
\def\@fpheader{\relax}
\makeatother

\title{On the structure of the large-$N$ expansion in SU($N$) Yang-Mills theory}

\author[a]{Marco Bochicchio,}
\author[b,a]{Mauro Papinutto,}
\author[b,a]{Francesco Scardino}

\affiliation[a]{Physics Department, INFN Roma1, 
	Piazzale A. Moro 2, Roma, I-00185, Italy}
\affiliation[b]{Physics Department, Sapienza University,Piazzale A. Moro 2, Roma, I-00185, Italy}

\emailAdd{marco.bochicchio@roma1.infn.it}
\emailAdd{mauro.papinutto@roma1.infn.it}
\emailAdd{francesco.scardino@roma1.infn.it}

\abstract{Recently, we have computed the short-distance asymptotics of the generating functional of Euclidean correlators of single-trace twist-$2$ operators in the large-$N$ expansion of SU($N$) Yang-Mills (YM) theory to the leading-nonplanar order. Remarkably, it has the structure of the logarithm of a functional determinant, but with the sign opposite to the one that would follow from the spin-statistics theorem for the glueballs. In order to solve this sign puzzle, we have reconsidered the proof in the literature that in the 't Hooft topological expansion of large-$N$ YM theory the leading-nonplanar contribution to the generating functional consists of the sum over punctures of $n$-punctured tori. We have discovered that for twist-$2$ operators it contains -- in addition to the $n$-punctured tori -- the normalization of tori with $1 \leq p \leq n$ pinches and $n-p$ punctures. Once the existence of the new sector is taken into account, the violation of the spin-statistics theorem disappears. Moreover, the new sector contributes trivially to the nonperturbative $S$ matrix because -- for example -- the $n$-pinched torus represents nonperturbatively a loop of $n$ glueball propagators with no external leg. This opens the way for an exact solution limited to the new sector that may be solvable thanks to the vanishing $S$ matrix.  }

\begin{document} 

		\maketitle
	\flushbottom

	\section{Introduction to large-$N$ YM theory and the sign puzzle}

It has been known for more than forty years that SU($N$)\footnote{To be precise, the original 't Hooft argument \cite{tHooft:1973alw} applies to the large-$N$ U($N$) (bare) theory. The SU($N$) case is the subject of the present paper.} Yang-Mills (YM) theory admits the 't Hooft large-$N$ topological expansion \cite{tHooft:1973alw} for the $n$-point connected correlators of gauge-invariant single-trace operators. To say it in a nutshell, the corresponding Feynman diagrams in 't Hooft double-line representation -- after a suitable gluing of reversely oriented lines -- are topologically classified \cite{tHooft:1973alw,Veneziano:1976wm} by the sum on the genus $g$ of $n$-punctured closed Riemann surfaces, where each topology is weighted by a factor $N^{\chi}$, with $\chi=2-2g-n$ the Euler characteristic of the Riemann surface. The 't Hooft topological expansion extends to large-$N$ QCD \cite{tHooft:1973alw,Veneziano:1976wm} -- and generally to large-$N$ gauge theories -- by eventually including punctured Riemann surfaces with boundaries.\par
It exactly matches the topology and weights \cite{tHooft:1973alw,Veneziano:1976wm} of a string theory \cite{Veneziano:1974dr}, with closed-string coupling $g_s=\frac{1}{N}$.
This matching has been advocated in favor of the existence \cite{tHooft:1973alw,Veneziano:1976wm,Aharony:1999ti} of a nonperturbative string solution of large-$N$ YM theory, QCD and, more generally, gauge theories, where -- in the asymptotically free case -- the dimensionful parameter of the string theory, conventionally referred to as the string tension, is identified \cite{Bochicchio:2017sgq} with the square of the renormalization-group (RG) invariant scale $\Lambda_{RG}$ of the gauge theory. In the would-be string solution of large-$N$ QCD glueballs are excitations of closed strings, mesons of open strings, and the aforementioned Riemann surfaces arise as their world-sheets -- the new ingredient of the canonical string solution \cite{Bochicchio:2017sgq} being the existence of a conformal field theory living on the string world-sheets that is employed to compute the $S$-matrix amplitudes \cite{Veneziano:1974dr} and possibly the correlators \cite{Gubser:1998bc} of the gauge theory. In fact, the existence of this would-be canonical string solution of (YM theory) QCD is fundamentally constrained \cite{Bochicchio:2017sgq} by the large-$N$ nonperturbative (non-)renormalization properties \cite{Bochicchio:2017teh} of $\Lambda_{(YM)QCD}$. \par
\begin{figure}[H]
	\centering
	\includegraphics[width=1\linewidth]{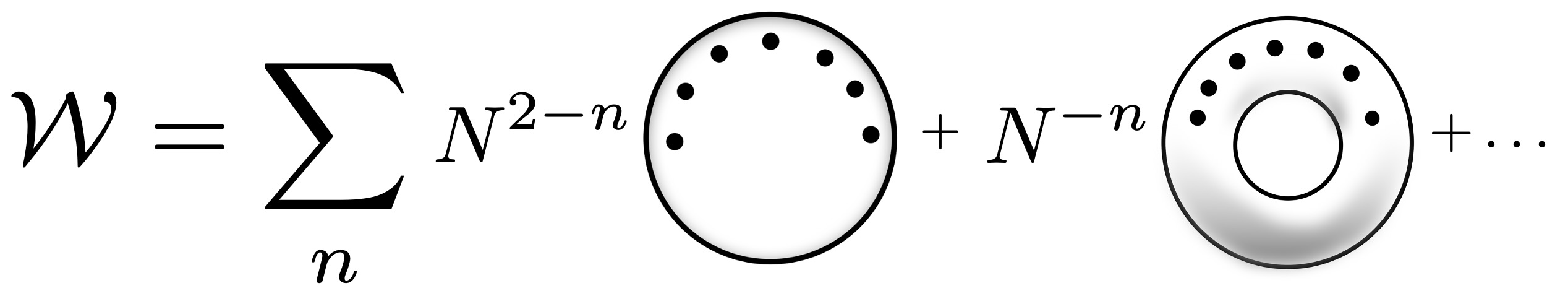}
	\caption{'t Hooft topological expansion of the generating functional, with $n$ the number of operator insertions and the ellipses higher genera.}
	\label{fig:spheretorus}
\end{figure}

By assuming the 't Hooft topological expansion, the generating functional $\mathcal{W}^E[J_{\mathcal O}]=\log \mathcal{Z}^E[J_{\mathcal{O}}]$ of Euclidean connected correlators of single-trace operators ${\mathcal O}$ in SU($N$) YM theory, with
\begin{align}
	&\mathcal{Z}^E[J_{\mathcal{O}}]=\frac{1}{\mathcal{Z}^E} \int \mathcal{D}A\, e^{-  S_{YM}+\sum_s\, \int J_{\mathcal{O}_s}\mathcal{O}_s}
\end{align}
reads (Fig. \refeq{fig:spheretorus})
\begin{equation}
	\label{thooft}
	\mathcal{W}^E[J_{\mathcal O}]=\mathcal{W}^E_{\text{sphere}}[J_{\mathcal O}]+\mathcal{W}^E_{\text{torus}}[J_{\mathcal O}]+ \cdots \,.
\end{equation}
Nonperturbatively, $\mathcal{W}^E_{\text{sphere}}[J_{\mathcal O}]$, which perturbatively is the ('t Hooft-)planar contribution \cite{tHooft:1973alw}, is a sum of tree diagrams involving glueball propagators and vertices, while $\mathcal{W}^E_{\text{torus}}[J_{\mathcal O}]$, which perturbatively is the leading-non('t Hooft-)planar contribution, is a sum of glueball one-loop diagrams. 
Nonperturbatively, $\mathcal{W}^E_{\text{torus}}[J_{\mathcal O}]$ should have the structure of the logarithm of a functional determinant \cite{Bochicchio:2016toi}. Indeed, in the yet-to-come nonperturbative solution of large-$N$ YM theory, the very same correlators should be computed by the correlators of a glueball field $\Phi$ with an infinite number of components, the corresponding generating functional being schematically \cite{Bochicchio:2016toi}
\begin{align}
	\mathcal{Z}^E_{\text{glueball}}[J] = &\frac{1}{\mathcal{Z}^E_{\text{glueball}}}\int\mathcal{D}\Phi \,e^{-S_{\text{glueball}}(\Phi)+\int \Phi \ast_1 J} \,,
\end{align}
with $S_{\text{glueball}}(\Phi)=\frac{1}{2}\int\,\Phi\ast_2(-\Delta+M^2)\Phi+\frac{1}{N}\frac{1}{3}\Phi \ast_3\Phi\ast_3\Phi+\cdots$,
where $\ast_2$ and $\ast_1$ are fixed below, the ellipses and $\ast_3$ respectively stand for $n$-glueball vertices with $n>3$ and some presently unknown operation on the glueball fields that, by assuming locality and Euclidean invariance, may involve derivatives. Hence, nonperturbatively the connected generating functional $\mathcal{W}^E_{\text{glueball}}[J] = \log\mathcal{Z}^E_{\text{glueball}}[J]$ reads to one loop of glueballs \cite{Bochicchio:2016toi}
\begin{align}
	\label{glueballW}
	&\mathcal{W}^E_{\text{glueball}}[J] = -S_{\text{glueball}}(\Phi_J)+\int \Phi_J \ast_1 J + \cdots \nonumber\\
	&-\frac{1}{2}\log\text{Det}\left(\ast_2(-\Delta+M^2)+\frac{1}{N}\ast_3\Phi_J\ast_3+\cdots\right) \,,
\end{align}
with $\Phi_J$ determined by $\frac{\delta S_{\text{glueball}}}{\delta\Phi}\Big\rvert_{\Phi_J}=\ast_1 J$. The minus sign in front of the $\log\Det$ in $\mathcal{W}^E_{\text{glueball}}[J] $ arises from the spin-statistics theorem \cite{streater2000pct}, since all the 
gauge-invariant glueball interpolating fields have integer spin, and thus the glueballs should be bosons.
The dictionary between $\mathcal{W}^E[J_{\mathcal O}]$ and $\mathcal{W}^E_{\text{glueball}}[J]$ is obtained by matching the corresponding
spectral representations -- as a sum of free propagators with residues $R_{sm}$ -- for the $2$-point correlators \cite{Migdal:1977nu} of $\mathcal{O}_s$ at  $N =+\infty$ that, by fixing $\ast_2$ according to the canonical normalization of the glueball kinetic term, uniquely determines the coupling of $J$ to the tower of glueball fields $\Phi \ast_1 J = \sum_{sm} \Phi_{sm} \sqrt{R_{sm}} J_{s}$.\par
Until recently nothing has been known quantitatively on $\mathcal{W}^E[J_{\mathcal O}]$ and $\mathcal{W}^E_{\text{glueball}}[J]$.
Actually, since the early days of large-$N$ QCD it has been known at a qualitative level that asymptotic freedom (AF) applies to large-$N$ correlators, and specifically to the nonperturbative $2$-point correlators \cite{Witten:1979kh,1987gauge}.\par
Accordingly, several years ago the ultraviolet (UV), i.e., short distance, constraints on the large-$N$ spectral representation of $2$-point correlators that follow from AF have been investigated at a quantitative level both for multiplicatively renormalizable operators \cite{Bochicchio:2013eda} and twist-$2$ operators that mix by renormalization \cite{Bochicchio:2023ols}.  \par
More recently, we have worked out the UV asymptotics of the generating functional $\mathcal{W}^E[J_{\mathcal O}]$ of Euclidean correlators of twist-$2$ operators in large-$N$ YM theory \cite{BPSpaper2} that follows from the RG, Callan-Symanzik equation and AF. 
For the aims of the present paper it suffices to report for simplicity just a particular case of our calculation (Appendices \ref{app:AA}, \ref{app:A}, \ref{app:B}, \ref{app:C}) that involves the nonchiral operators of twist-$2$ with even spin and maximal-spin component in the $p_+$ direction in Minkowskian space-time.
In the light-cone gauge $A_+=0$ they read \cite{BPS1,BPSpaper2}
\begin{equation}\label{eq:Os}
	\mathbb{O}_{s} =\frac{1}{2N}\bar{A}^a(x) \mathcal{Y}_{s-2}^{\frac{5}{2}}(\overrightarrow{\partial}_+,\overleftarrow{\partial}_+) A^a(x) \,,
\end{equation}
where $s=2,4,6,\ldots$, the sum over repeated color indices, $a=1,2,\ldots N^2-1$, is understood, and
\begin{align}
	\label{defY}
	\mathcal{Y}_{s-2}^{\frac{5}{2}}(\overrightarrow{\partial}_+,\overleftarrow{\partial}_+) =\overleftarrow{\partial}_+ (i\overrightarrow{\partial}_++i\overleftarrow{\partial}_+)^{s-2}C^{\frac{5}{2}}_{s-2}\left(\frac{\overrightarrow{\partial}_+-\overleftarrow{\partial}_+}{\overrightarrow{\partial}_++\overleftarrow{\partial}_+}\right)\overrightarrow{\partial}_+ \,,
\end{align}
with $C^{\frac{5}{2}}_{s-2}$ Gegenbauer polynomials (Appendix \ref{app:A}). Our calculation has been performed in two steps that we briefly recall.
First, we have computed to the lowest perturbative order
-- directly from its functional-integral definition as a Gaussian integral (Appendix \ref{app:A})
\begin{align}
	\mathcal{Z}_{\text{conf}}[J_{\mathbb{O}}]=\frac{1}{Z} \int \mathcal{D}A\mathcal{D}\bar{A}\, e^{ \int -i \bar{A}^a \square A^a+\sum_s\, J_{\mathbb{O}_s}\mathbb{O}_s \, d^4x}
\end{align}
the generating functional of the connected conformal correlators in Minkowskian space-time \cite{BPSpaper2} 
\begin{align}
	\label{Wconfsign} 
	\mathcal{W}_{\text{conf}}[J_{\mathbb{O}}]=-(N^2-1)\log\text{Det}\left(\mathbb{I}+\frac{1}{2}i\square^{-1} \frac{J_{\mathbb{O}_s}}{N}\otimes
	\mathcal{Y}_{s-2}^{\frac{5}{2}}\right)
\end{align}
and -- by analytical continuation -- the corresponding object in Euclidean space-time (Appendix \ref{app:B})
\begin{align}
	\label{WEconfsign}
	\mathcal{W}^E_{\text{conf}}[J_{\mathbb{O}^E}]=-(N^2-1)\log\text{Det}\left(\mathbb{I}+\frac{1}{2}\Delta^{-1}\frac{J_{\mathbb{O}^E_{s}}}{N}\otimes\mathcal{Y}_{s-2}^{E\frac{5}{2}}\right) \,,
\end{align}
where $\mathbb{I}$ is the identity in space-time, the sum over repeated $s$ indices is understood and $\mathcal{Y}_{s-2}^{E\frac{5}{2}}$ is obtained from $\mathcal{Y}_{s-2}^{\frac{5}{2}}$ by the substitution $\partial_+ \rightarrow i \partial_z$ \cite{BPS1,BPSpaper2} (Appendix \ref{app:B}). \par
Second (Appendix \ref{app:C}), by means of a careful choice of the renormalization scheme (Appendix \ref{app:C1}) that reduces the mixing of the above operators to the multiplicatively renormalizable case (Appendix \ref{app:C2}) to all orders of perturbation theory \cite{Bochicchio:2021geometry,BPSpaper2}, we have lifted $\mathcal{W}^E_{\text{conf}}[J_{\mathcal O}]$ to the generating functional of the RG-improved asymptotic correlators $\mathcal{W}^E_{\text{asym}}[J_{\mathcal O},\lambda]$, as all the coordinates are uniformly rescaled by a factor $\lambda \rightarrow 0$, that inherits the very same structure of the logarithm of a functional determinant (Appendix \ref{app:C4})
\begin{align}
	\label{Wasymsign}
	\mathcal{W}^E_{\text{asym}}[J_{\mathbb{O}^E},\lambda]=-(N^2-1)\log\text{Det}\left(\mathbb{I}+\frac{1}{2}\frac{Z_{\mathbb{O}_s}(\lambda)}{\lambda^{s+2}}\Delta^{-1}\frac{J_{\mathbb{O}^E_{s}}}{N}\otimes\mathcal{Y}_{s-2}^{E\frac{5}{2}}\right) \,,
\end{align}
$Z_{\mathbb{O}_s}$ being one-loop exact (Appendix \ref{app:C2}) in the above scheme \cite{Bochicchio:2021geometry} and independent of $N$ \cite{Bochicchio:2021geometry,BPSpaper2} (Appendix \ref{app:C3})
\begin{eqnarray} \label{Z}
	Z_{\mathbb{O}_s}(\lambda) = \Bigg(\frac{g(\mu)}{g(\frac{\mu}{\lambda})}\Bigg)^{\frac{\gamma_{0\mathbb{O}_s}}{\beta_0}} \,.
\end{eqnarray}
The conformal generating functional admits the large-$N$ expansion that follows trivially from Eq. \eqref{Wconfsign}
\begin{equation}
	\mathcal{W}_{\text{conf}}[J_{\mathcal O}]=\mathcal{W}_{\text{conf\,sphere}}[J_{\mathcal O}]+\mathcal{W}_{\text{conf\,torus}}[J_{\mathcal O}]
\end{equation}
and the asymptotic generating functional as well
\begin{equation}
	\mathcal{W}^E_{\text{asym}}[J_{\mathcal O},\lambda]=\mathcal{W}^E_{\text{asym\,sphere}}[J_{\mathcal O},\lambda]+\mathcal{W}^E_{\text{asym\,torus}}[J_{\mathcal O},\lambda] \,,
\end{equation}
with
\begin{align}
	\mathcal{W}^E_{\text{asym\,torus}}[J_{\mathbb{O}^E},\lambda]= +\log\text{Det}\left(\mathbb{I}+\frac{1}{2}\frac{Z_{\mathbb{O}_s}(\lambda)}{\lambda^{s+2}}\Delta^{-1} \frac{J_{\mathbb{O}^E_s}}{N}\otimes\mathcal{Y}_{s-2}^{E\frac{5}{2}}\right) \,.
\end{align}
Remarkably, the above equation reproduces the $\log\Det$ structure of the glueball one-loop generating functional in Eq. \eqref{glueballW}, which it should be asymptotic to in the UV thanks to the AF, but with the sign opposite to the one that follows from the spin-statistics theorem for the glueballs. Nevertheless, we verify the correctness of the above sign, which is inherited from Eqs. \eqref{Wconfsign}, \eqref{WEconfsign} and \eqref{Wasymsign}, according to the obvious sign due to the spin-statistics theorem for gluons in one-loop perturbation theory, and the fact that in the SU($N$) theory there are exactly $N^2-1$ gluons.\par

\section{Spin-statistics theorem and refinement of 't Hooft topological expansion}

The aim of the present paper is to solve the sign puzzle and to discuss the implications of its solution. 
Essentially, there are two possible way-outs: Either the spin-statistics theorem is violated or the 't Hooft topological expansion (Fig. 1) and the corresponding effective action in Eq. \eqref{glueballW} need considerable refinements for the correlators of twist-$2$ operators. As we shall show momentarily, only the second alternative applies in large-$N$ YM theory. \par
\begin{figure}[H]
	\centering
	\includegraphics[width=0.8\linewidth]{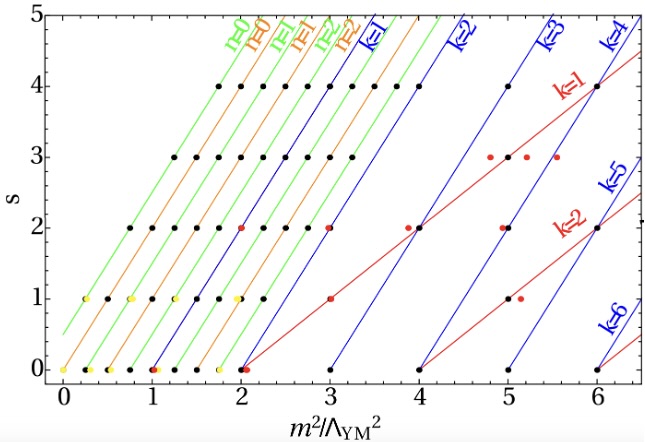}
	\caption{Glueball (blue and red) and meson (green and orange) Regge trajectories \cite{bochicchio2016glueball}.}
	\label{fig:screenshot001}
\end{figure}

In fact, the spin-statistics theorem applies in full generality only to theories with a finite number of local fields \cite{harish1947infinite,feldman1966unitarity,streater1967local,Casalbuoni:2006fa}.
If the number of fields is infinite, there exist rigorous counterexamples \cite{harish1947infinite,feldman1966unitarity,streater1967local} to the spin-statistics theorem.
They involve massive infinite dimensional representations of the Lorentz group that are the relevant ones in YM theory because of its mass gap \cite{jaffe2006quantum}. Yet, they have infinite mass degeneracy, since they decompose \cite{feldman1966unitarity,streater1967local}, according to the Wigner theorem \cite{wigner1939unitary}, into the sum of irreducible representations of the Poincar\'e group corresponding to an infinite number of particles of any spin, all having the same mass.
In large-$N$ YM theory this would imply the existence of vertical Regge trajectories that is hardly acceptable both theoretically and numerically because of contrary evidence from analysis \cite{bochicchio2016glueball} of lattice calculations (Fig. \ref{fig:screenshot001}).\par
\begin{figure}[H]
	\centering
	\includegraphics[width=1\linewidth]{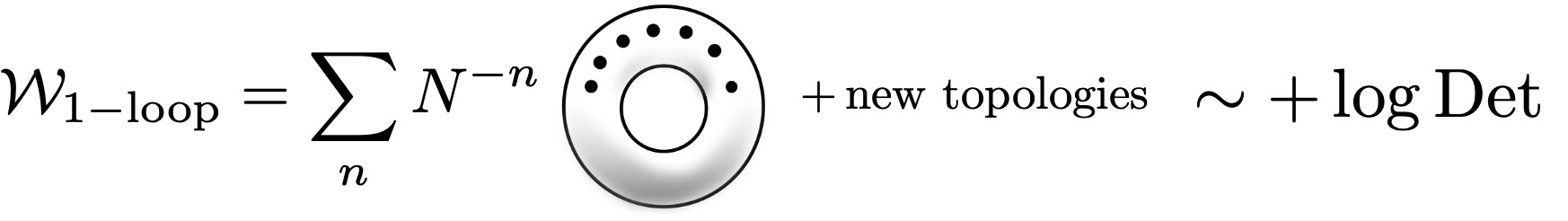}
	\caption{New topologies may solve the sign puzzle compatibly with the spin-statistics theorem.}
	\label{fig:newtopo}
\end{figure}

\begin{figure}[H]
	\centering
	\includegraphics[width=0.6\linewidth]{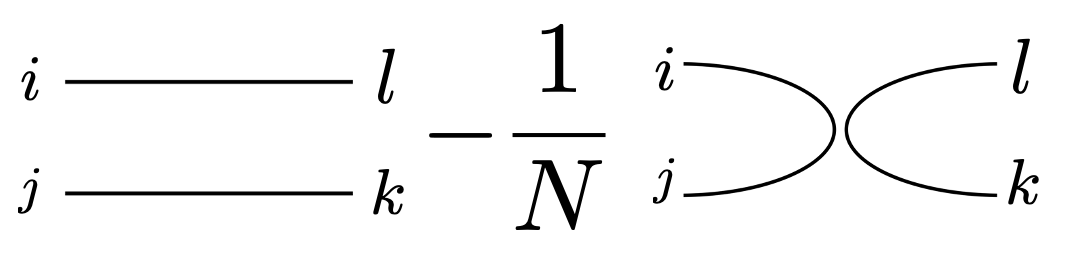}
	\caption{Leading and subleading SU($N$) gluon propagator in double line.}
	\label{fig:propagator}
\end{figure}

Hence, it remains the second alternative (Fig. \ref{fig:newtopo}).
In order to understand why new topologies arise in the generating functional of correlators of twist-$2$ operators, we have reconsidered the proof of the 't Hooft topological expansion in the U($N$) versus SU($N$) YM theory.
The delicate point of the proof in the SU($N$) theory is that -- contrary to the original U($N$) case \cite{tHooft:1973alw} -- the gluon propagator has a leading and subleading $\frac{1}{N}$ contribution in its double-line representation \cite{marino2015instantons,maltoni2003color} (Fig. \ref{fig:propagator}):
$\langle A^i_{\, j} A^k_{\, l}\rangle \propto I^{i \,  k}_{\, j \, l}-P^{i \, k}_{\, j \, l}$, with $A^i_{\, j}=A^a(\lambda^a)^i_{\, j}$, $\lambda^a$ in the fundamental representation, $\Tr(\lambda^a \lambda^b)=\frac{1}{2}\delta^{ab}$, $[\lambda^a, \lambda^b]=if^{abc}\lambda^c$, $I^{i \, k}_{\, j\, l}=\delta^i_{\, l}\delta^k_{\, j}$ the components of the identity $I$ in the u($N$) Lie algebra \footnote{In double-line notation the product of two matrices, $B$ and $C$, is defined by $(BC)^{i \, m}_{\, j \, n}=B^{i \, k}_{\, j \, l}C^{l \, m}_{\, k \, n}$, with the color trace $\Tr B = B^{i \, j}_{\, j \, i}$},
$P^{i \, k}_{\, j \, l}=\frac{1}{N}\delta^i_{\, j}\delta^k_{\, l}$ the components of the u($1$) projector $P$, and $i,l,j,k=1, \ldots, N$.\par
\begin{figure}[H]
	\centering
	\includegraphics[width=1\linewidth]{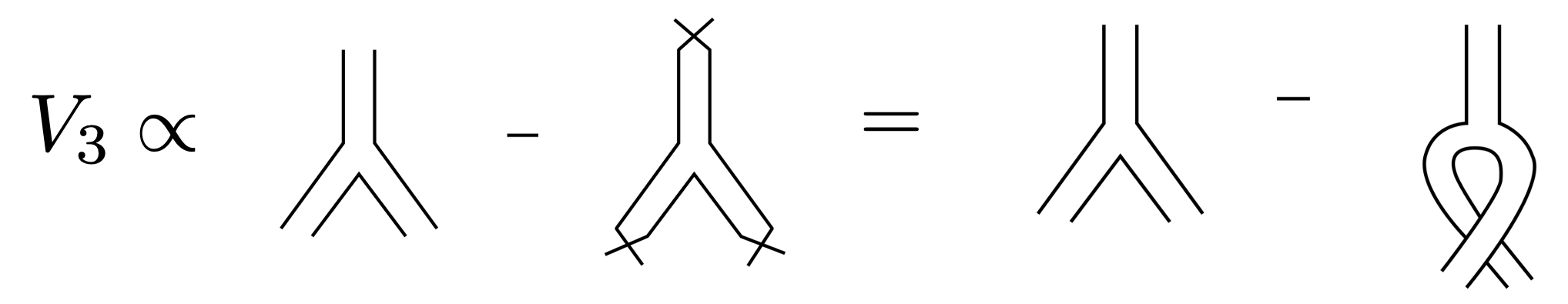}
	\caption{$3$-gluon action vertex in double line \cite{marino2015instantons}.}
	\label{fig:vertex}
\end{figure}

\begin{figure}[H]
	\centering
	\includegraphics[width=0.7\linewidth]{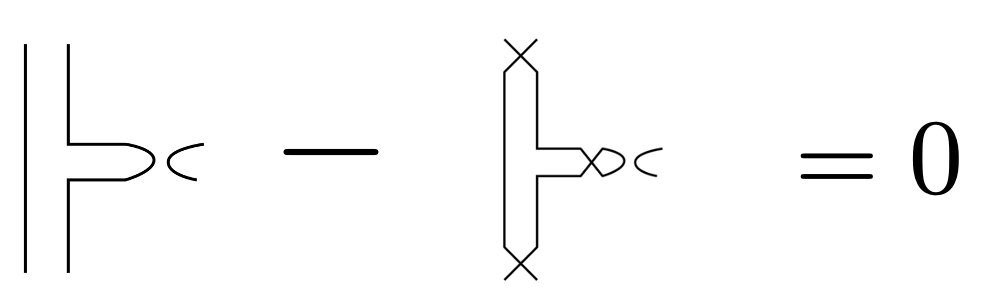}
	\caption{Decoupling of the subleading propagator from the $3$-gluon action vertex.}
	\label{fig:proof}
\end{figure}

Yet, the subleading propagator, if it is attached to the vertices of the action,
$V_3 \propto \frac{g}{\sqrt{N}}f^{abc}$ (Fig. \ref{fig:vertex}) and 
$V_4 \propto \frac{g^2}{N}f^{abe}f^{ecd}$, does not contribute \cite{marino2015instantons,maltoni2003color}.
This is rather obvious \cite{maltoni2003color}, since the subleading propagator is a u($1$) contribution, while the action vertices are purely nonabelian. This fact has a simple graphical proof  (Fig. \ref{fig:proof}).
Therefore, in the SU($N$) theory the topology of connected vacuum diagrams containing vertex insertions matches the 't Hooft topological expansion with weights $N^\chi$, where the gluon propagator is just the leading one  \cite{marino2015instantons,maltoni2003color} in Fig. \ref{fig:propagator}.\par
At this point, generalizing the structure of $V_3$ and $V_4$, we observe that the above argument extends to connected correlators of single-trace operators provided that the corresponding local vertices are proportional to a matrix product \cite{maltoni2003color} of $(T^a)^b_c= -if^{abc}$, $(T^{a_2} \dots T^{a_{n-1}})^{a_1}_{a_n}=2\Tr(\lambda^{a_1}[\lambda^{a_2}, \dots [\lambda^{a_{n-1}},\lambda^{a_{n}}] \dots])$ that requires at least $3$-gluon operators in perturbation theory.  \par
Indeed, the above vertices in double line have no $\frac{1}{N}$ correction \cite{maltoni2003color} and their contraction with the subleading propagator vanishes because they involve nested commutators. Therefore, there is a vast class of single-trace operators which the 't Hooft topological expansion applies to in SU($N$) YM theory \footnote{From the above proof it follows that this is a very special property of the aforementioned operators. In fact, it does not hold true for generic operators, for which a case-by-case analysis is required \cite{BPS3}.}.\par
\begin{figure}[H]
	\centering
	\includegraphics[width=0.4\linewidth]{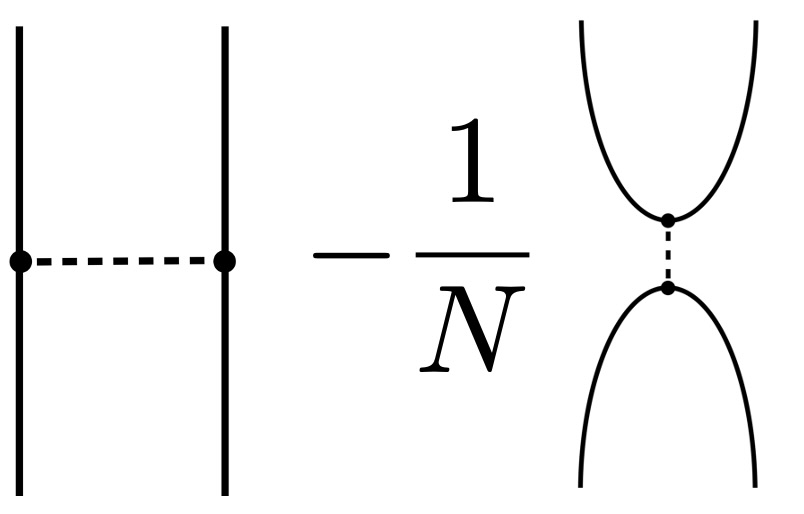}
	\caption{Local vertex for $2$-gluon operators in double-line. The dotted lines identify the punctures.}
	\label{fig:2vertex}
\end{figure}

Yet, we point out that the above proof does not apply to $2$-gluon operators, and specifically to twist-$2$ operators. Indeed, in this case -- even in the adjoint representation -- their local vertex involves $\delta^{ab}$, as opposed to $f^{abc}$, and contains a $\frac{1}{N}$ correction (Fig. \ref{fig:2vertex}), exactly as the gluon propagator does. 
In fact, it has been suggested to consider the U($N$) \cite{tHooft:1973alw,marino2015instantons,Aharony:1999ti} -- rather than the SU($N$) -- theory, where perturbatively no subleading correction both to the gluon propagator and (bare) $2$-gluon vertices occur -- though gauge-invariant single-trace twist-$2$ operators have not been specifically discussed. \par
Nevertheless, this seemingly easy way-out is pointless for asymptotically free theories that allow us to resum the (renormalized) RG-improved generating functional into the logarithm of a functional determinant
\begin{align} \label{15}
	\mathcal{W}^E_{{\text{U}(N)}\text{asym}}[J_{\mathbb{O}},\lambda] =&-(N^2-1)\log\Det\left(\mathbb{I}+\frac{1}{2}\frac{Z_{\mathbb{O}_s}(\lambda)}{\lambda^{s+2}}\Delta^{-1}\frac{J_{\mathbb{O}^E_{s}}}{N}\otimes\mathcal{Y}_{s-2}^{E\frac{5}{2}}\right)\nonumber\\
	&-\log\Det\left(\mathbb{I}+\frac{1}{2}\frac{1}{\lambda^{s+2}}\Delta^{-1}\frac{J_{\mathbb{O}^E_{s}}}{N}\otimes\mathcal{Y}_{s-2}^{E\frac{5}{2}}\right) \,,
\end{align}
while the U($1$) contribution in the second line decouples and is exactly conformal and not only asymptotic because the U($1$) theory is free. 
Indeed, asymptotically in the UV the above result is dominated by only the SU($N$) contribution for the correlators of all the twist-$2$ operators but the stress-energy tensor, since for them $\gamma_{0\mathbb{O}_s}$ in Eq. \eqref{Z} is positive \cite{Belitsky:1998gc,Belitsky:2004sc,BPSpaper2} (Appendix \ref{app:C3}). \par
\begin{figure}[H]
	\centering
	\includegraphics[width=1\linewidth]{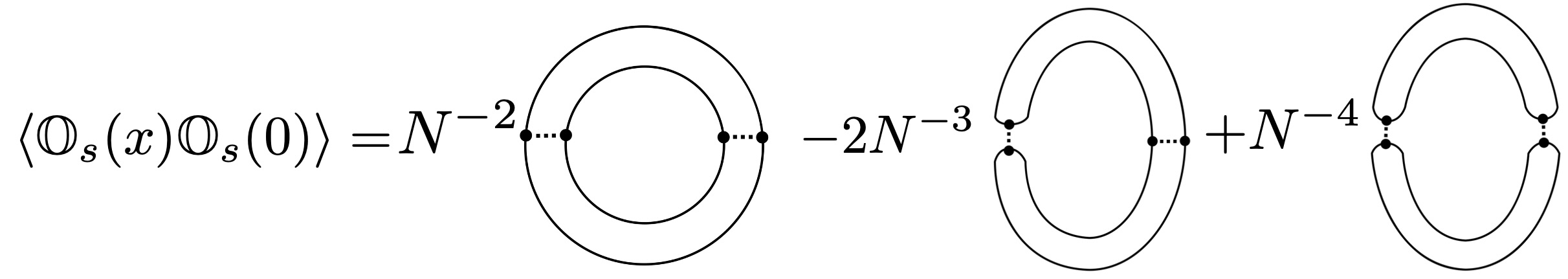}
	\caption{$2$-point correlators of twist-$2$ operators to the leading perturbative order, with signs and weights as in Fig. \ref{fig:nonrenorm1}.}
	\label{fig:2pointcolor}
\end{figure}
\begin{figure}[H]
	\centering
	\includegraphics[width=1\linewidth]{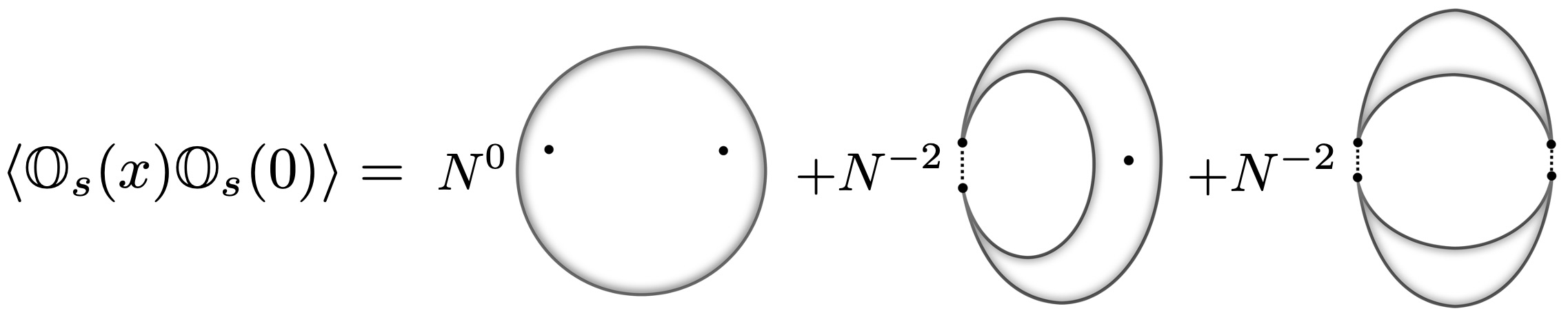}
	\caption{Topology and weights of diagrams in Fig. \ref{fig:2pointcolor}.}
	\label{fig:2pointnewtopos}
\end{figure}
\begin{figure}[H]
	\centering
	\includegraphics[width=1\linewidth]{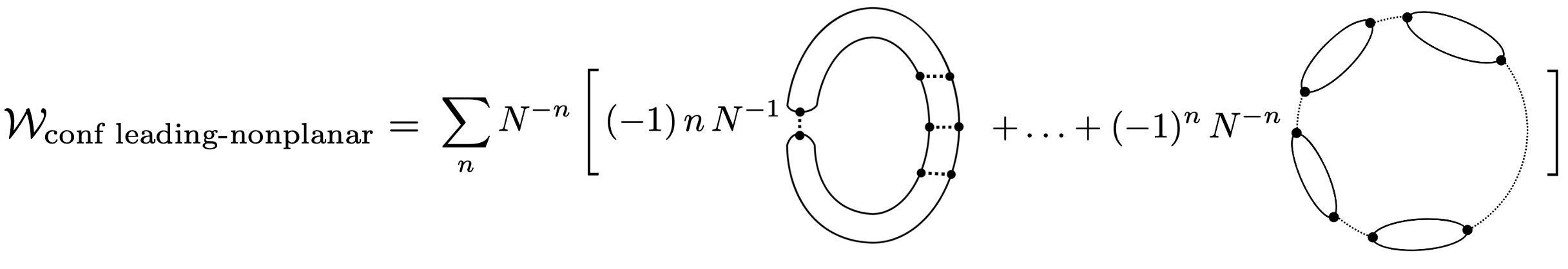}
	\caption{The refined perturbative expansion of the conformal leading-nonplanar generating functional: The square brackets contain the sum over $1 \leq p \leq n$ of diagrams that are the normalization of pinched annuli involving $n-p$ pairs of punctures and $p$ pinches arising from $p$ insertions of $P$, which give rise to the weights $N^{-p}$ compensated for by the color traces represented by loops. Some relative signs and combinatorics in the expansion of the second line of Eq. \eqref{Wconf1} with respect to the corresponding planar contributions in the first line are also displayed.}
	\label{fig:nonrenorm1}
\end{figure}

Hence, the sign problem for the U($N$) theory is exactly the same as for the SU($N$) theory \footnote{For correlators only involving the stress-energy tensor, Eq. \eqref{15} does not seem to imply any sign problem because in this specific case the leading asymptotics of the $O(N^0)$ SU($N$) and U(1) contributions actually cancel.
	But this is an illusion, since the asymptotics of Eq.(15) is sufficient for the issue about the spin-statistics theorem to arise,  but by no means necessary: In fact, since both perturbative and nonperturbative renormalization inevitably separate the $O(N^0)$ SU($N$) and U(1) sectors, the issue with the spin-statistics theorem is apparent for the stress-energy correlators as well in the SU($N$) sector -- where a running coupling and mass gap occur, as opposed to the decoupled free conformal U(1) sector -- of the renormalized U($N$) theory.
	In any case, the new topologies that we shall introduce momentarily in the main text, whose existence in the SU($N$) theory is actually independent of the issue about the spin-statistics theorem but instrumental to solve it, also occur for the correlators of the stress-energy tensor both in the SU($N$) theory and, by the above argument, in the renormalized U($N$) one. It is only in the bare U($N$) theory that no new topologies occur, according to the original 't Hooft argument.}. 
To solve the sign problem in the SU($N$) theory, we rewrite Eq. \eqref{WEconfsign} identically as
\begin{align}
	\label{Wconf0}
	\mathcal{W}^E_{\text{conf}}[J_{\mathbb{O}^E}] =-\log\Det\left(\mathcal{I}+\frac{1}{2}(I-P)\Delta^{-1}\frac{J_{\mathbb{O}^E_{s}}}{N}\otimes\mathcal{Y}_{s-2}^{E\frac{5}{2}}\right) \,,
\end{align}
where the term containing $I-P$ involves the SU($N$) propagator (Fig. \ref{fig:propagator}) and $\mathcal{I}$ is the identity in both color and space-time.
The aforementioned equality follows by noticing that the color trace $\Tr$ in the loop expansion of Eq. \eqref{Wconf0} containing the insertion of $n$ sources $J_{\mathbb{O}^E_{s}}$ produces the overall factor 
\begin{align}
	\Tr(I-P)^n= \Tr(I-P)=N^2-1
\end{align}
that occurs in Eq. \eqref{WEconfsign}.
In order to keep the 't Hooft double-line representation that only involves the leading propagator also beyond the planar limit of the SU($N$) theory, we transfer in Eq. \eqref{Wconf0} the $\frac{1}{N}$ dependence from the propagator to the local vertex for $2$-gluon operators (Fig. \ref{fig:2vertex}) by the identity
\begin{align}
	\label{Wconf01}
	&\mathcal{W}^E_{\text{conf}}[J_{\mathbb{O}^E}] \nonumber\\
	&=-\log\Det\left(\mathcal{I}+\frac{1}{2}(I-P)\Delta^{-1}\frac{J_{\mathbb{O}^E_{s}}}{N}\otimes\mathcal{Y}_{s-2}^{E\frac{5}{2}}\right)\nonumber\\
	&=-\log\Det\left(\mathcal{I}+\frac{1}{2}I\Delta^{-1}\frac{J_{\mathbb{O}^E_{s}}}{N}\otimes\mathcal{Y}_{s-2}^{E\frac{5}{2}}-\frac{1}{2}P\Delta^{-1}\frac{J_{\mathbb{O}^E_{s}}}{N}\otimes\mathcal{Y}_{s-2}^{E\frac{5}{2}}\right) \nonumber \\
	&=-\log\Det\Bigg[\Big(\mathcal{I}+\frac{1}{2}\Delta^{-1} I \frac{J_{\mathbb{O}^E_{s}}}{N}\otimes\mathcal{Y}_{s-2}^{E\frac{5}{2}}\Big) \nonumber \\
	&\,\,\,\,\,\,\,\,\, \left(\mathcal{I}-\frac{1}{2}\Big(\mathcal{I}+\frac{1}{2}\Delta^{-1} I \frac{J_{\mathbb{O}^E_{s}}}{N}\otimes\mathcal{Y}_{s-2}^{E\frac{5}{2}}\Big)^{-1}\Delta^{-1}P\frac{J_{\mathbb{O}^E_{s}}}{N}\otimes\mathcal{Y}_{s-2}^{E\frac{5}{2}}\right)\Bigg] \,,
\end{align}
so that
\begin{align}
	\label{Wconf1}
	\mathcal{W}^E_{\text{conf}}[J_{\mathbb{O}^E}]
	&=-\log\Det\left(\mathcal{I}+\frac{1}{2}\Delta^{-1} I \frac{J_{\mathbb{O}^E_{s}}}{N}\otimes\mathcal{Y}_{s-2}^{E\frac{5}{2}}\right)\nonumber\\
	&\quad-\log\Det\Bigg(\mathcal{I}-\frac{1}{2}\Big(\mathcal{I}+\frac{1}{2}\Delta^{-1} I \frac{J_{\mathbb{O}^E_{s}}}{N}\otimes\mathcal{Y}_{s-2}^{E\frac{5}{2}}\Big)^{-1} \Delta^{-1}P\frac{J_{\mathbb{O}^E_{s}}}{N}\otimes\mathcal{Y}_{s-2}^{E\frac{5}{2}}\Bigg) \,. \nonumber\\
\end{align}
The first $\log \Det$ above is the ('t Hooft-)planar contribution -- involving the leading local vertex (Fig. \ref{fig:2vertex}) -- and the second $\log \Det$ the leading-nonplanar one -- involving at least one subleading local vertex (Fig. \ref{fig:2vertex}) carrying the factor of $P$.
The latter gives rise to new topologies. For example, for $2$-point correlators (Fig. \ref{fig:2pointcolor}),
by the standard 't Hooft gluing of reversely oriented lines, the first diagram in Fig. \ref{fig:2pointcolor} -- that is the planar contribution -- leads to a $2$-punctured sphere and the remaining ones -- by representing a $2$-punctured disk as an infinite strip and gluing the opposite edges of the strip to get an infinite cylinder, i.e., a $2$-punctured sphere (Fig. \ref{fig:2pointnewtopos}) -- to possibly disconnected punctured spheres with at least two punctures pairwise identified (Fig. \ref{fig:2pointnewtopos}). 
By generalizing the new diagrams in Fig. \ref{fig:2pointcolor}, the leading-nonplanar contributions due to the second $\log \Det$ in Eq. \eqref{Wconf1} involve the new topological sector (Fig. \ref{fig:nonrenorm1}).
After the RG improvement, the leading-nonplanar asymptotic generating functional reads
\begin{align}
	\label{Wasnp}
	&\mathcal{W}^E_{\text{asym nonplanar}}[J_{\mathbb{O}^E},\lambda] \nonumber\\
	&=-\log\Det\Bigg(\mathcal{I}-\frac{1}{2}\Big(\mathcal{I}+ \frac{1}{2} \frac{Z_{\mathbb{O}_s}(\lambda)}{\lambda^{s+2}}  \Delta^{-1} I \frac{J_{\mathbb{O}^E_{s}}}{N}\otimes\mathcal{Y}_{s-2}^{E\frac{5}{2}}\Big)^{-1} \frac{Z_{\mathbb{O}_s}(\lambda)}{\lambda^{s+2}}\Delta^{-1}P\frac{J_{\mathbb{O}^E_{s}}}{N}\otimes\mathcal{Y}_{s-2}^{E\frac{5}{2}}\Bigg)\nonumber \\
	&=+\log\text{Det}\left(\mathbb{I}+\frac{1}{2}\frac{Z_{\mathbb{O}_s}(\lambda)}{\lambda^{s+2}}\Delta^{-1} \frac{J_{\mathbb{O}^E_s}}{N}\otimes\mathcal{Y}_{s-2}^{E\frac{5}{2}}\right) \,.
\end{align}
Some observations are in order.\par
\begin{figure}[H]
	\centering
	\includegraphics[width=1\linewidth]{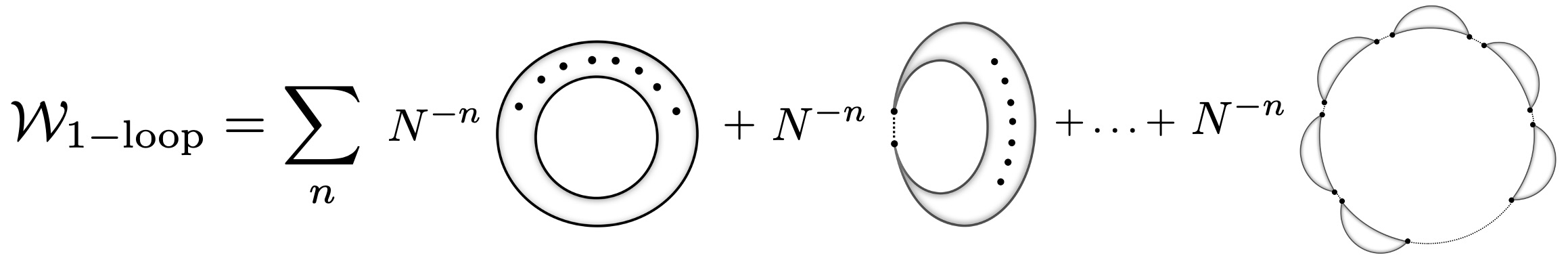}
	\caption{The refined topological expansion of the glueball one-loop generating functional.}
	\label{fig:w1loop}
\end{figure}
\begin{figure}[H]
	\centering
	\includegraphics[width=1\linewidth]{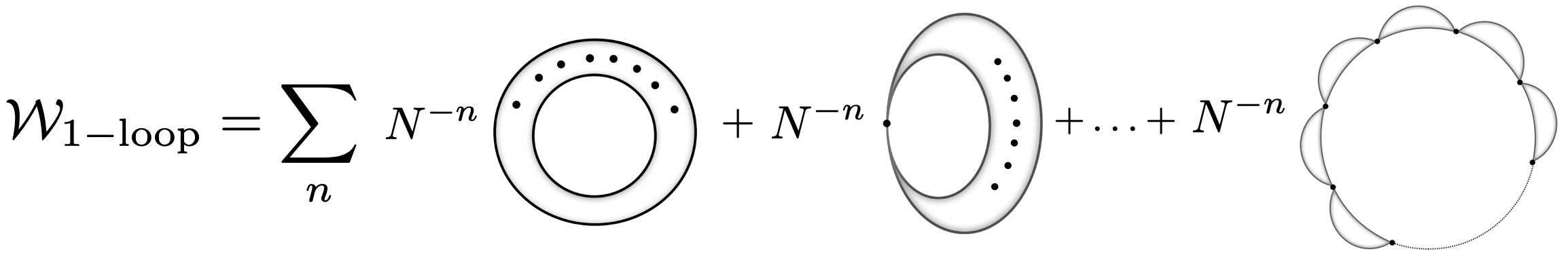}
	\caption{The new topologies in Fig. \ref{fig:w1loop} arise from the components of the normalization of $p$-pinched tori with $n-p$ punctures obtained by cutting the pinches.}
	\label{fig:w2loop}
\end{figure}

First, we can think of the leading-nonplanar RG-improved generating functional above as the resummation by the Callan-Symanzik equation (Appendix \ref{app:C}) of the leading logarithms from both pinched and smooth tori (Figs. \ref{fig:w1loop} and \ref{fig:w2loop}), the pinched tori arising from a suitable gluing of the strips 
in Fig. \ref{fig:nonrenorm1} for each $n$, analogously to Fig. \ref{fig:2pointnewtopos} for $n=2$. \par
Indeed, though the punctured smooth tori are suppressed in perturbation theory with respect to the pinched ones because the smooth tori inevitably involve the action vertices that carry powers of the 't Hooft gauge coupling,
in the bare theory their leading divergent parts must be proportional, according to the fundamental principles of renormalizable perturbation theory, to the lowest-order contribution furnished by the pinched tori. \par
For example, because of multiplicative renormalizability and scaling symmetry \footnote{Since $\gamma_{0\mathbb{O}_{s}}$ is diagonal, scaling symmetry also holds to order $g^2$ in schemes different from the conformal one, which instead is necessary to implement invariance under (collinear) conformal inversions \cite{Braun:2003rp}.} the bare $2$-point correlators of twist-2 operators read in $4-2\epsilon$ dimensions to order $g^2$ in perturbation theory 
\begin{align}
	\langle\mathbb{O}_{Bs}(x) \mathbb{O}_{Bs}(y)\rangle=&\left(1-\frac{1}{N^{2}}\right)C_s \frac{(x-y)_{+}^{2s}}{(|x-y|^{2})^{2s+2}} \nonumber \\
	&\biggl[1- \left(\frac{ \gamma_{0\mathbb{O}_{s}}g^2}{\epsilon}\right) +  \gamma_{0\mathbb{O}_{s}} g^2 \log\frac{1}{|x-y|^{2}\mu^{2}}+ \cdots \biggr]
\end{align}
and similarly for the analytically continued correlators to Euclidean space-time. \par
Therefore, in the bare theory the $2$-punctured smooth tori produce perturbative divergences
\begin{align}
	\langle\mathbb{O}_{Bs}(x) \mathbb{O}_{Bs}(y)\rangle_{\text{smooth torus}}=&-\frac{1}{N^{2}} C_s \frac{(x-y)_{+}^{2s}}{(|x-y|^{2})^{2s+2}} \nonumber \\
	&\biggl[- \left(\frac{\gamma_{0\mathbb{O}_{s}}g^2}{\epsilon}\right) +  \gamma_{0\mathbb{O}_{s}} g^2 \log\frac{1}{|x-y|^{2}\mu^{2}}+ \cdots \biggr]
\end{align}
that, after summing with the contributions in Fig. \ref{fig:2pointcolor} corresponding to the pinched tori in Fig. \ref{fig:2pointnewtopos}
\begin{align}
	\langle\mathbb{O}^B_{s}(x) \mathbb{O}^B_{s}(y)\rangle_{\text{pinched tori}}=&\left(-\frac{2}{N^{2}}+\frac{1}{N^{2}}\right) C_s \frac{(x-y)_{+}^{2s}}{(|x-y|^{2})^{2s+2}} \nonumber \\
	&=-\frac{1}{N^{2}} C_s \frac{(x-y)_{+}^{2s}}{(|x-y|^{2})^{2s+2}} \, ,
\end{align}
are cancelled by means of multiplicative renormalization (Appendix \ref{app:C3})
\begin{equation}
	Z_{ss}^2=1+\frac{\gamma_{0\mathbb{O}_{s}} g^2}{ \epsilon} + \cdots
\end{equation}
by counterterms that have precisely the space-time structure of pinched tori
\begin{align}
	-\left(\frac{ \gamma_{0\mathbb{O}_{s}}g^2}{\epsilon}\right)\frac{1}{N^{2}} C_s \frac{(x-y)_{+}^{2s}}{(|x-y|^{2})^{2s+2}}
\end{align}
To say it in a nutshell, the smooth tori mix with the pinched ones by perturbative renormalization, so that their contribution cannot be separated in the RG-improved generating functional in Eq. \eqref{Wasnp}. \par
Second, as a consequence, the leading-nonplanar renormalized asymptotic RG-improved correlators obtained by the resummation of the leading logarithms have precisely the structure of the pinched correlators -- i.e., of the lowest-order conformal result -- multiplied by the renormalized factors due to the anomalous dimensions and beta function provided by      
the smooth tori according to Eq. \eqref{Wasnp}.\par
Finally, and most remarkably, now the overall sign of the first $\log \Det$ in Eq. \eqref{Wasnp} is consistent with the bosonic statistics for the glueballs, but at the price of 
introducing a refined topological expansion where, in addition to the $n$-punctured tori, punctured spheres (Fig. \ref{fig:w1loop}) that are the normalization \cite{Chan2021ModuliSO,Witten:2012ga} of $p$-pinched and $n-p$-punctured tori (Fig. \ref{fig:w2loop}) arise, with $1 \leq p \leq n$. 
Interestingly, in the new sector the weights differ \footnote{Indeed, adding one pinch increases $\chi$ by one with respect to the corresponding smooth surface \cite{pinch}.} from $N^{\chi}$, unless the pinched tori are thought of as a specific degeneration of the smooth ones, where some cycles of the smooth tori collapse to points and collide with an equal number of punctures, as we explain in the following.

\section{Nonperturbative interpretation of the refined topological expansion}

We provide now a nonperturbative interpretation of our refined topological expansion for twist-$2$ operators -- that incidentally explains nonperturbatively the above mixing -- in terms of the effective theory of glueballs that is closely related  -- though not exactly coincident -- to the dual-graph representation of Riemann surfaces \cite{Chan2021ModuliSO,Witten:2012ga} recalled below.\par
\begin{figure}[H]
	\centering
	\includegraphics[width=0.7\linewidth]{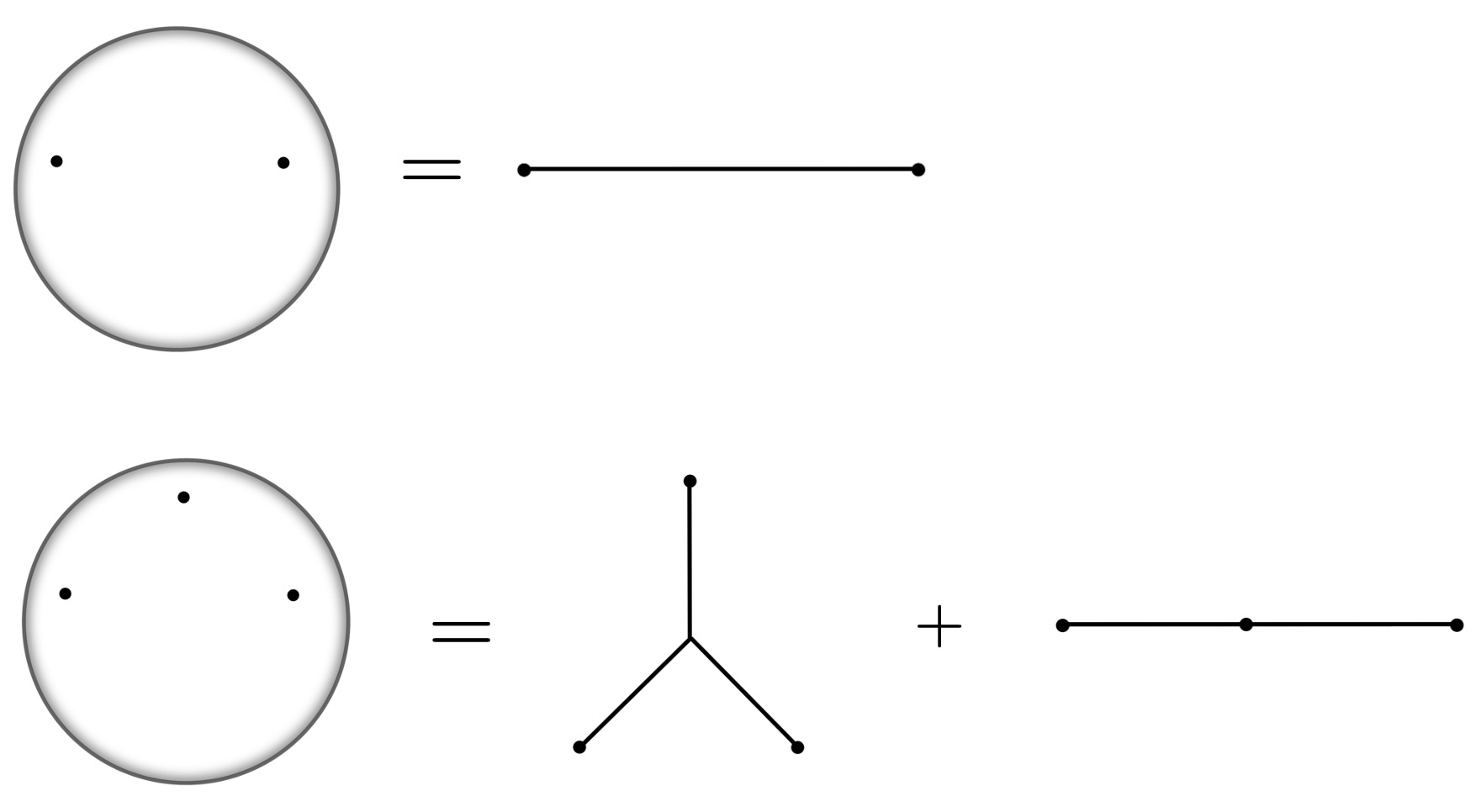}
	\caption{Glueball tree graphs dual to punctured spheres in the effective theory of glueballs.}
	\label{fig:spheres}
\end{figure}

It has been known for more than forty years that, in the effective theory of glueballs, punctured spheres correspond to glueball tree graphs \cite{Migdal:1977nu,Witten:1979kh}. 
Specifically, the sphere with two punctures corresponds to an infinite sum of glueball propagators \cite{Migdal:1977nu}, while the sphere with three punctures corresponds in Minkowskian space-time to vertices that are sums of three or two glueball poles \cite{Witten:1979kh} (Fig. \ref{fig:spheres}).\par
\begin{figure}[H]
	\centering
	\includegraphics[width=1\linewidth]{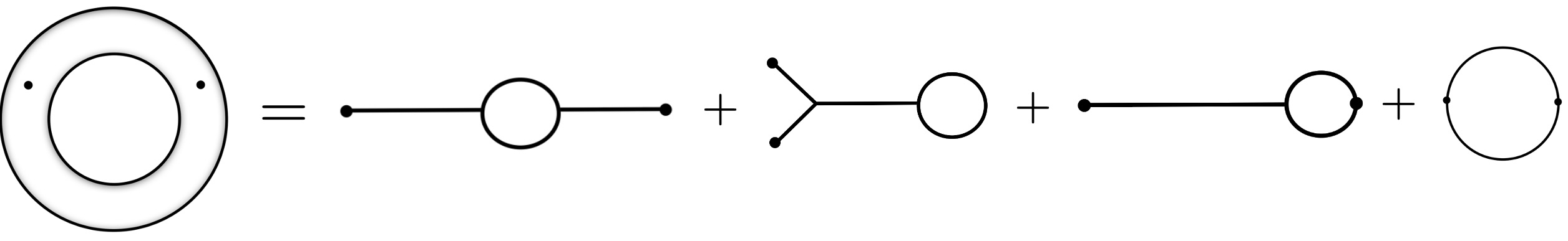}
	\caption{Glueball one-loop graphs dual to the $2$-punctured torus.}
	\label{fig:torus}
\end{figure}
\begin{figure}[H]
	\centering
	\includegraphics[width=0.7\linewidth]{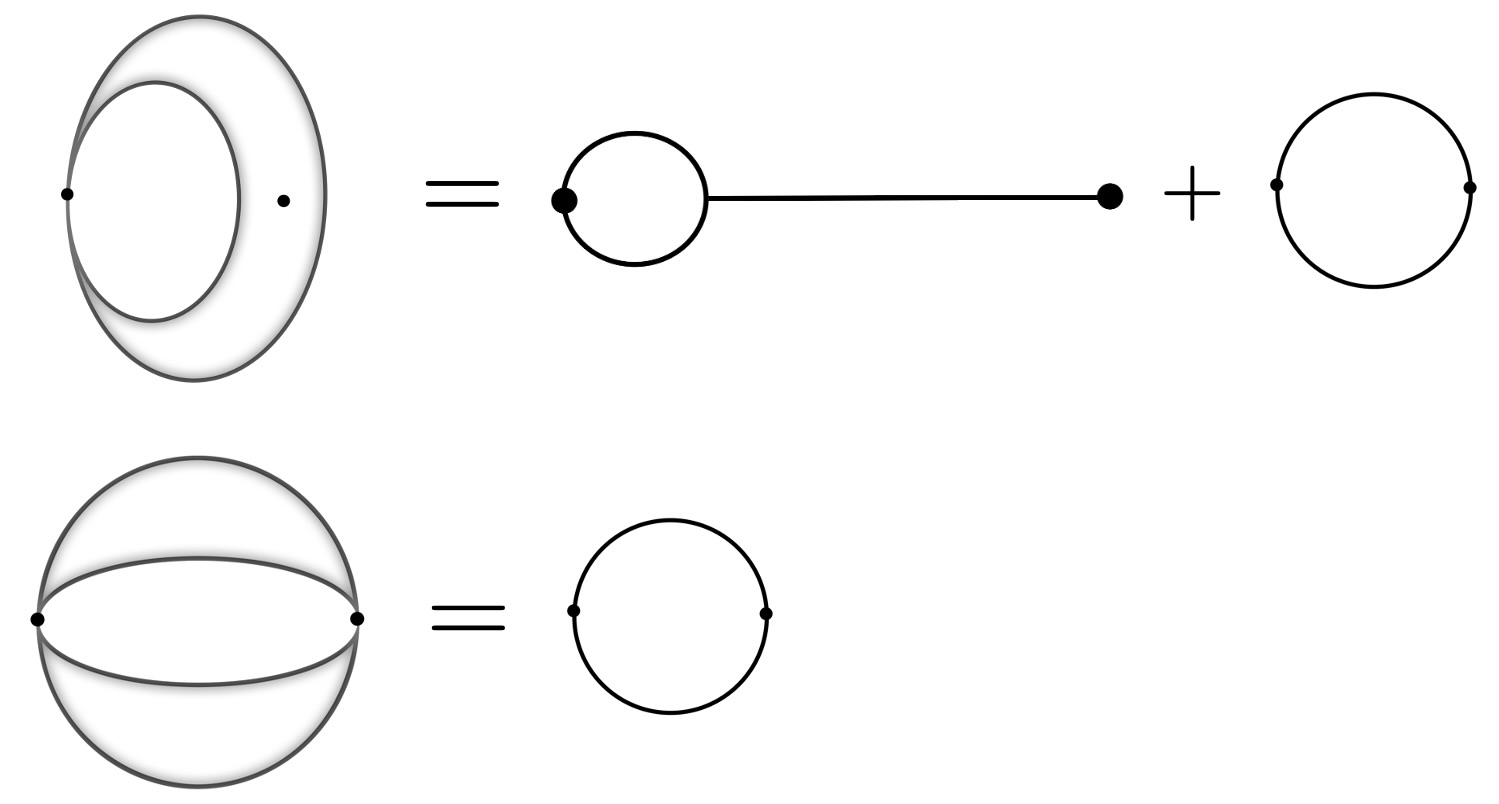}
	\caption{Glueball one-loop graphs dual to pinched tori.}
	\label{fig:maxdisc}
\end{figure}
\begin{figure}[H]
	\centering
	\includegraphics[width=1\linewidth]{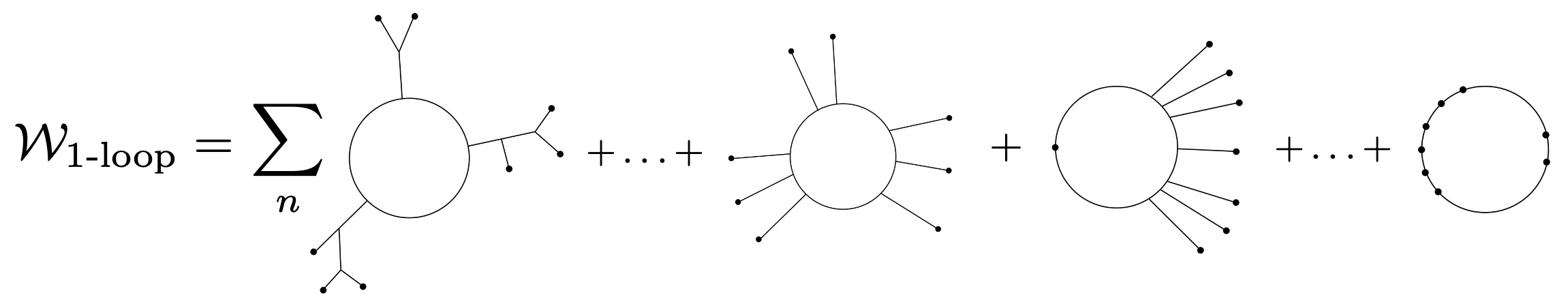}
	\caption{Glueball one-loop generating functional.}
	\label{fig:w1loopspectral}
\end{figure}

For simplicity, in the following we skip $n$-glueball vertices with $n>3$ as in Eq. \eqref{glueballW}.
Incidentally, we point out that the last graph in Fig. \ref{fig:spheres} contributes zero to the $S$ matrix because of the missing glueball external leg.
Analogously, the $2$-punctured torus may contribute, in addition to glueball one-loop two-leg graphs, also graphs with only one or zero external glueball legs (Fig. \ref{fig:torus})
and, similarly, our new topologies may contribute the graphs in Fig. \ref{fig:maxdisc}. Incidentally, the partial matching of the graphs in Fig. \ref{fig:maxdisc} with the ones in Fig. \ref{fig:torus} nonperturbatively explains the aforementioned mixing by renormalization. Analogous statements hold for the graphs in Figs. \ref{fig:w2loop} and \ref{fig:w1loopspectral}.\par
Both graphs in Fig. \ref{fig:maxdisc} contribute zero to the $S$ matrix as well. More generally, all the new topologies contribute zero to the $S$ matrix, since they miss at least one external glueball leg, as for the last two graphs in Fig. \ref{fig:w1loopspectral}.
It is an open problem -- though -- whether an effective action of the kind in Eq. \eqref{glueballW} exists that also captures nonperturbatively the contribution of the entire new topological sector for the correlators. \par
However, independently of the existence of the above effective action, for the graphs that only contain bivalent vertices, i.e., the maximally pinched ones in Fig. \eqref{fig:w2loop}, which are in one-to-one correspondence with diagrams only involving the subleading vertex and no leading vertex in Eq. \eqref{Wasnp}, we verify the spin-statistics theorem directly by means of our asymptotic computation
\begin{align}
	\label{Wasnp1}
	\mathcal{W}^E_{\text{asym maximally pinched}}[J_{\mathbb{O}^E}, \lambda] 
	&=-\log\Det\Bigg(\mathcal{I} - \frac{1}{2} \frac{Z_{\mathbb{O}_s}(\lambda)}{\lambda^{s+2}}\Delta^{-1}P\frac{J_{\mathbb{O}^E_{s}}}{N}\otimes\mathcal{Y}_{s-2}^{E\frac{5}{2}}\Bigg)\nonumber \\
	&=-\log\text{Det}\left(\mathbb{I}-\frac{1}{2}\frac{Z_{\mathbb{O}_s}(\lambda)}{\lambda^{s+2}}\Delta^{-1} \frac{J_{\mathbb{O}^E_s}}{N}\otimes\mathcal{Y}_{s-2}^{E\frac{5}{2}}\right)
\end{align}
thanks to the minus sign in the last line above.\par
Finally, the occurrence of pinched tori with punctures resembles the Deligne-Mumford (DM) compactification \cite{Chan2021ModuliSO,Witten:2012ga} of the moduli space of punctured closed Riemann surfaces that arises in canonical string theories \cite{Witten:2012ga} due to their underlying conformal structure  \cite{Witten:2012ga}.\par
Yet, in general our pinched tori do not occur in the DM compactification of $n$-punctured tori, whose components only involve punctured closed Riemann surfaces with $\chi < 0$  \cite{Chan2021ModuliSO} and, in particular, spheres with at least three punctures \cite{Chan2021ModuliSO,Witten:2012ga} -- because in the DM compactification punctures and pinches never collide \cite{Chan2021ModuliSO,Witten:2012ga}, according to the conformal structure of canonical string theories \cite{Witten:2012ga} --  while the components of our pinched tori may contain $2$-punctured spheres with $\chi = 0$. \par
Indeed, contrary to the DM compactification \cite{Chan2021ModuliSO}, the dual graphs to our pinched tori may contain bivalent vertices as in the last graphs of Figs. \ref{fig:maxdisc} and \ref{fig:w1loopspectral}. As a consequence, no canonical closed-string theory that admits the DM compactification may contain the new topological sector.

\section{Conclusions}

In the large-$N$ SU($N$) YM theory a new topological sector exists that refines the 't Hooft topological expansion for the correlators of twist-$2$ operators, both perturbatively and nonperturbatively. The new topologies solve the sign puzzle, specifically in the maximally pinched sector. \par
The new topological sector dominates the UV asymptotics of the correlators of twist-$2$ operators, but contributes zero to the nonperturbative $S$ matrix, since nonperturbatively it consists of tori with at least one pinch, corresponding to glueball one-loop diagrams with at least one glueball external leg missing. \par
As in general the new topologies do not arise from the DM compactification of the moduli space of punctured tori, no canonical string theory admitting it may exist for the correlators of large-$N$ YM theory in the new topological sector, but it may exist for the $S$-matrix amplitudes.\par
Finally, the existence of the new topological sector -- specifically, the maximally pinched one -- that contributes zero to the $S$ matrix, but nontrivially to the correlators, opens the way for a nonperturbative solution limited to the new sector by a topological field/string theory -- noncanonical in the sense of the present paper -- along the lines foreseen in \cite{Bochicchio:2016toi} \footnote{By reinterpreting the coupling to D-branes in \cite{Bochicchio:2016toi} as the generating functional of correlators in the new sector, rather than of the collinear $S$-matrix amplitudes \cite{MBarxiv}.}.

\section*{Acknowledgments}
We would like to thank Tommaso Bechini for working out the wonderful plots of Riemann surfaces.
\newpage
\appendix

\section{YM functional integral in the light-cone gauge} \label{app:AA}

To make the present paper self-contained, we report the computation of the Euclidean generating functional $\mathcal{W}^E_{\text{asym}}[J_{\mathbb{O}}]$ -- limited to the twist-$2$ operators in the main text -- in the coordinate representation of the RG-improved asymptotic correlators of all the collinear twist-$2$ operators \cite{BPSpaper2}. 
We also fix some notations and conventions. \par
The Minkowskian YM action is
\begin{equation}
	S_{YM} = -\frac{1}{2}\int d^4x\,\Tr F_{\mu\nu}F^{\mu\nu} \,,
\end{equation}
where
\begin{align}
	F_{\mu\nu} = \partial_\mu A_\nu - \partial_\nu A_\mu +i\frac{g}{\sqrt{N}}\left[A_\mu,A_\nu\right]
\end{align}
and
\begin{equation}
	A_\mu= A_\mu^a T^a \,,
\end{equation}
with the Hermitian generators of the SU($N$) Lie algebra
\begin{equation}
	\left[T^a,T^b\right] = i f^{abc}T^c
\end{equation}
in the fundamental representation normalized as
\begin{equation}
	\Tr(T^aT^b) = \frac{\delta^{ab}}{2}
\end{equation}
and $g^2 = Ng^2_{YM}$ the 't Hooft coupling \cite{tHooft:1973alw}. We set
\begin{align}
	&V_+ = \frac{V_0+ V_3}{\sqrt{2}}\qquad V_{-} = \frac{V_0- V_3}{\sqrt{2}}\nonumber\\
	&V = \frac{V_1+i V_2}{\sqrt{2}}\qquad \bar{V} = \frac{V_1-i V_2}{\sqrt{2}}
\end{align}
for a vector $V_{\mu}$.
We choose the light-cone gauge
\begin{equation}
	A_+ = 0 \,.
\end{equation}
After integrating out $A_-$, the YM action in the light-cone gauge reads \cite{Belitsky:2004sc}
\begin{align}
	S_{YM}(A, \bar A) = -\int& \bar{A}^a \square A^a+2\frac{g}{\sqrt{N}} f^{abc}(A^a\partial_+\bar{A}^b\bar{\partial}\partial_+^{-1}A^c+\bar{A}^a\partial_+A^b\partial\partial_+^{-1}\bar{A}^c) \nonumber\\
	&+2\frac{g^2}{N}f^{abc}f^{ade}\partial_+^{-1}(A^b\partial_+ \bar{A}^c)\partial_+^{-1}(\bar{A}^d\partial_+A^e)  \,d^4x \,.
\end{align} 
The v.e.v. of a product of local gauge-invariant operators $\mathcal{O}_i(A) $ that do not depend on $A_{-}$ reads
\begin{align}
	&\langle\mathcal{O}_1(x_1)\ldots \mathcal{O}_n(x_n)\rangle=\frac{1}{Z}\int \mathcal{D}A\mathcal{D}\bar{A}\, e^{iS_{YM}(A,\bar A)} \mathcal{O}_1(x_1)\ldots \mathcal{O}_n(x_n) \,.
\end{align}

\section{Minkowskian conformal generating functional $\mathcal{W}_{\text{conf}}$ \label{app:A}}

To the leading perturbative order -- where YM theory is conformal \cite{Braun:2003rp} -- it reduces to
\begin{align}
	\langle \mathcal{O}_1(x_1)\ldots \mathcal{O}_n(x_n)\rangle=\frac{1}{Z}\int \mathcal{D}A\mathcal{D}\bar{A}\, e^{-i \int d^4x\, \bar{A}^a \square A^a}\mathcal{O}_1(x_1)\ldots \mathcal{O}_n(x_n) \,,
\end{align}
with
\begin{equation}
	\square= g^{\mu\nu}\partial_{\mu}\partial_{\nu}=\partial_0^2-\sum_{i=1}^{3}\partial_i^2 \,,
\end{equation}
where we employ the mostly minus metric $g^{\mu\nu}$ in Minkowskian space-time \cite{BPS1}. 
The even-spin balanced \footnote{We refer to either balanced or unbalanced operators, if in their spinorial representation they carry either an equal or different number of dotted and undotted indices, respectively.} conformal collinear twist-$2$ operators \cite{BPS1} read (Eq. \eqref{eq:Os})
\begin{align} \label{CO}
	&\mathbb{O}_{s} =\frac{1}{2N}\bar{A}^a(x) \mathcal{Y}_{s-2}^{\frac{5}{2}}(\overrightarrow{\partial}_+,\overleftarrow{\partial}_+) A^a(x)\ \,,
\end{align}
where $s=2,4,6,\ldots$, the sum over repeated color indices is understood, and (Eq. \eqref{defY})
\begin{align}\label{gegen}
	\mathcal{Y}_{s-2}^{\frac{5}{2}}(\overrightarrow{\partial}_+,\overleftarrow{\partial}_+) 
	&= \overleftarrow{\partial}_+ (i\overrightarrow{\partial}_++i\overleftarrow{\partial}_+)^{s-2}C^{\frac{5}{2}}_{s-2}\left(\frac{\overrightarrow{\partial}_+-\overleftarrow{\partial}_+}{\overrightarrow{\partial}_++\overleftarrow{\partial}_+}\right)\overrightarrow{\partial}_+\nonumber\\
	&=\frac{\Gamma(3)\Gamma(s+3)}{\Gamma(5)\Gamma(s+1)}i^{s-2}\sum_{k=0}^{s-2} {s\choose k}{s\choose k+2}(-1)^{s-k} \overleftarrow{\partial}_{+}^{s-k-1} \overrightarrow{\partial}_{+}^{k+1} \,.
\end{align}
In the present paper the above operators are already normalized so that their $2$-point correlators are of order $1$ for large $N$.
The corresponding generating functional of conformal correlators to the leading order reads
\begin{align}
	\mathcal{Z}_{\text{conf}}[J_{\mathbb{O}}]=\frac{1}{Z} \int \mathcal{D}A\mathcal{D}\bar{A}\, e^{-i \int d^4x\, \bar{A}^a \square A^a}\exp\left(\int d^4x\sum_s\, J_{\mathbb{O}_s}\mathbb{O}_s\right) \,,
\end{align}
where the currents $J_{\mathbb{O}_s}$ are defined to be zero for $s$ different from the spin of the corresponding operator. Explicitly,
\begin{align}
	\mathcal{Z}_{\text{conf}}[J_{\mathbb{O}}]
	= \frac{1}{Z}\int &\mathcal{D}A\mathcal{D}\bar{A}\, e^{-i\int d^4x\, \bar{A}^a \square A^a}\nonumber\\
	&\exp\Bigg(\frac{1}{2}\int d^4x\sum_s\,\frac{J_{\mathbb{O}_{s}}}{N}\bar{A}^a(x) \mathcal{Y}_{s-2}^{\frac{5}{2}}(\overrightarrow{\partial}_+,\overleftarrow{\partial}_+) A^a(x)\Bigg) \,.
\end{align}
The above functional integral is quadratic in the elementary fields. Therefore, it may be computed exactly as a Gaussian integral \cite{BPSpaper2}
\begin{equation}
	\mathcal{Z}_{\text{conf}}[J_{\mathbb{O}}] =\Det^{-1}( \mathcal{M}) \,,
\end{equation}
where 
\begin{align}
	\label{matrixM}
	&\mathcal{M}^{ab}=\delta^{ab} \Big( i\square-\frac{1}{2}\sum_s\frac{J_{\mathbb{O}_{s}}}{N}\otimes\mathcal{Y}_{s-2}^{\frac{5}{2}} \Big) 
\end{align}
and the symbol $\otimes$ implies that the right and left derivatives do not act on the sources $J$.
Hence, the generating functional of the connected conformal correlators reads \cite{BPSpaper2}
\begin{align}
	\label{Wgen1}
	\mathcal{W}_{\text{conf}}[J_{\mathbb{O}}]  &= 	\log\mathcal{Z}_{\text{conf}}[J_{\mathbb{O}}] \nonumber\\ &=-\log\Det\left(\mathcal{I}+\frac{1}{2}i\square^{-1}\frac{J_{\mathbb{O}_{s}}}{N}\otimes\mathcal{Y}_{s-2}^{\frac{5}{2}}\right) \,,
\end{align}
where $\mathcal{I}$ is the identity in both color and space-time and the sum over repeated spin indices is understood. 
Substituting Eq. \eqref{gegen} and performing the color trace, we explicitly obtain
{\small
	\begin{align}
		\label{wexplicit}
		&{\mathcal{W}_{\text{conf}}}[J_{\mathbb{O}}]\nonumber\\
		&=-(N^2-1)\log\Det\left(\mathbb{I}+\frac{1}{2}i\square^{-1}\frac{J_{\mathbb{O}_{s}}}{N}\otimes\mathcal{Y}_{s-2}^{\frac{5}{2}}\right)\nonumber\\
		&= -(N^2-1)\log\Det\left(\mathbb{I}+\frac{1}{2}\frac{\Gamma(3)\Gamma(s+3)}{\Gamma(5)\Gamma(s+1)}\sum_{k=0}^{s-2}{s\choose k}{s\choose k+2}(i\overrightarrow{\partial}_+)^{s-k-1}i \square^{-1}\frac{J_{\mathbb{O}_{s}}}{N}(i\overrightarrow{\partial}_+)^{k+1} \right) \,,
	\end{align}
}
where $\mathbb{I}$ is the identity in space-time. To get the last line in Eq. \eqref{wexplicit} we have employed the definition in Eq. \eqref{gegen} and
\begin{align}
	i\square^{-1}\overleftarrow{\partial}_{+}^{s-k-1} = (-1)^{s-k-1}\overrightarrow{\partial}_{+}^{s-k-1} i\square^{-1}
\end{align}
that follows from (minus) the propagator in the coordinate representation \cite{BPS1}
\begin{equation}
	\label{propagator}
	\frac{1}{4\pi^2}\frac{1}{\rvert x-y \rvert^2-i\epsilon} =i \square^{-1}(x-y) \,.
\end{equation}

\section{Euclidean conformal generating functional $\mathcal{W}^E_{\text{conf}}$ \label{app:B}}	

The Minkowskian correlators can be analytically continued to Euclidean space-time by the Wick rotation \cite{BPS1}
\begin{align} 
	\label{Wick}
	&x^+= \frac{x^0 + x^3}{\sqrt{2}}
	\rightarrow -ix^z= -i \frac{x^4+ix^3}{\sqrt{2}}
\end{align}
that implies
\begin{equation} \label{bc}
	\frac{1}{\rvert x \rvert^2-i\epsilon}\rightarrow-\frac{1}{x^2}\,.
\end{equation}
The analytically continued operators read
\begin{equation}
	\mathbb{O}_s \rightarrow (-1)^{s+1}\Tr \partial_z \bar{A}^{E}(\overrightarrow{\partial_z} + \overleftarrow{\partial_z})^{s-2}C^{\frac{5}{2}}_{s-2}\Bigg(\frac{\overrightarrow{\partial_z} - \overleftarrow{\partial_z}}{\overrightarrow{\partial_z}+\overleftarrow{\partial_z}}\Bigg)\partial_z A^{E} = \mathbb{O}^{A\,E}_s \,,
\end{equation}
with $\partial_z= \frac{\partial}{\partial x^z}$.
Therefore, Eq. \eqref{Wgen1} becomes \cite{BPSpaper2}
\begin{align}
	\mathcal{W}^E_{\text{conf}}[J_{\mathbb{O}^E}]  &= 	\log\mathcal{Z}^E_{\text{conf}}[J_{\mathbb{O}^E}] \nonumber\\ &=-\log\Det\left(\mathcal{I}+\frac{1}{2}\Laplace^{-1}\frac{J_{\mathbb{O}^E_{s}}}{N}\otimes\mathcal{Y}_{s-2}^{E\,\frac{5}{2}}\right) \,,
\end{align}
with
\begin{equation}
	\Laplace= \delta_{\mu\nu}\partial_{\mu}\partial_{\nu}=\partial_4^2+\sum_{i=1}^{3}\partial_i^2 
\end{equation}
and
\begin{equation} \label{Laplacian}
	\Laplace^{-1}(x-y)=-	\frac{1}{4\pi^2}\frac{1}{(x-y)^2} \,.
\end{equation}
Performing the color trace, we get
{\small
	\begin{align}
		\label{WgenE1}
		&\mathcal{W}^E_{\text{conf}}[J_{\mathbb{O}^E}] =-(N^2-1)\log\Det\left(\mathbb{I}+\frac{1}{2}\Laplace^{-1}\frac{J_{\mathbb{O}^E_{s}}}{N}\otimes\mathcal{Y}_{s-2}^{E\,\frac{5}{2}}\right)\nonumber\\
		&=-(N^2-1)\log\Det\left(I+\frac{1}{2}\frac{\Gamma(3)\Gamma(s+3)}{\Gamma(5)\Gamma(s+1)}\sum_{k=0}^{s-2}{s\choose k}{s\choose k+2}(-\overrightarrow{\partial}_z)^{s-k-1}\Laplace^{-1}\frac{J_{\mathbb{O}^E_{s}}}{N}(-\overrightarrow{\partial}_z)^{k+1} \right) \,.
	\end{align}
}

\section{Euclidean RG-improved asymptotic correlators}  \label{app:C}

\subsection{Operator mixing  \label{app:C1}}

We briefly summarize the construction of the RG-improved asymptotic correlators following \cite{BPSpaper2,Bochicchio:2021geometry}.
The renormalized Euclidean correlators
\begin{equation}\label{key}
	\langle \mathcal{O}_{k_1}(x_1)\ldots \mathcal{O}_{k_n}(x_n) \rangle = G^{(n)}_{k_1 \ldots k_n}( x_1,\ldots,  x_n; \mu, g(\mu))
\end{equation}
satisfy the Callan-Symanzik equation
\begin{align}\label{CallanSymanzik}
	& \Big(\sum_{\alpha = 1}^n x_\alpha \cdot \frac{\partial}{\partial x_\alpha} + \beta(g)\frac{\partial}{\partial g} + \sum_{\alpha = 1}^n D_{\mathcal{O}_\alpha}\Big)G^{(n)}_{k_1 \ldots k_n} + \sum_a \Big(\gamma_{k_1a}(g) G^{(n)}_{ak_2 \ldots k_n} \nonumber\\
	&\quad+ \gamma_{k_2a}(g) G^{(n)}_{k_1 a k_3 \ldots k_n} \cdots +\gamma_{k_n a}(g) G^{(n)}_{k_1 \ldots a}\Big) = 0\,,
\end{align}
with solution
\begin{align}\label{csformula}
	&G^{(n)}_{k_1 \ldots k_n}(\lambda x_1,\ldots, \lambda x_n; \mu, g(\mu)) \nonumber \\
	&= \sum_{j_1 \ldots j_n} Z_{k_1 j_1} (\lambda)\ldots Z_{k_n j_n}(\lambda)\hspace{0.1cm} \lambda^{-\sum_{i=1}^nD_{\mathcal{O}_{j_i}}} G^{(n)}_{j_1 \ldots j_n }( x_1, \ldots, x_n; \mu, g(\frac{\mu}{\lambda}))\,,
\end{align}
where $D_{\mathcal{O}_i}$ is the canonical dimension of $\mathcal{O}_i(x)$ and $\gamma(g)=\gamma_0 g^2+ \cdots$ the matrix of the anomalous dimensions, with
\begin{equation}\label{eqZ}
	\Bigg(\frac{\partial}{\partial g} + \frac{\gamma(g)}{\beta(g)}\Bigg)Z(\lambda) = 0
\end{equation}
in matrix notation, and
\begin{equation} \label{ZZ}
	Z(\lambda) = P\exp \Big(\int_{g(\mu)}^{g(\frac{\mu}{\lambda})}\frac{\gamma(g')}{\beta(g')} dg'\Big)\,.
\end{equation}
Eq. \eqref{csformula} greatly simplifies if a renormalization scheme exists where $Z(\lambda)$ is diagonalizable to all orders of perturbation theory, and specifically one-loop exact, with eigenvalues $Z_{\mathcal{O}_i}(\lambda)$ \cite{Bochicchio:2021geometry}
\begin{equation} \label{zz}
	Z_{\mathcal{O}_i}(\lambda) = \Bigg(\frac{g(\mu)}{g(\frac{\mu}{\lambda})}\Bigg)^{\frac{\gamma_{0\mathcal{O}_i}}{\beta_0}} \,,
\end{equation}
where $\gamma_{0\mathcal{O}_i}$ are the eigenvalues of $\gamma_0$. In the above scheme Eq. \eqref{csformula} contains only one term 
\begin{align}\label{csformuladiag}
	&G^{(n)}_{j_1 \ldots j_n}(\lambda x_1,\ldots, \lambda x_n; \mu, g(\mu)) \nonumber \\
	&=Z_{\mathcal{O}_{j_1}}(\lambda) \ldots Z_{\mathcal{O}_{j_n}}(\lambda)\,\lambda^{-\sum_{i=1}^nD_{\mathcal{O}_{j_i}}}G^{(n)}_{j_1 \ldots j_n }( x_1, \ldots, x_n; \mu, g(\frac{\mu}{\lambda}))\,.
\end{align}
Then, as $\lambda \rightarrow 0$, in any renormalization scheme, the renormalized correlator in the right-hand side above admits the perturbative asymptotic expansion in terms of the renormalized coupling $g(\frac{\mu}{\lambda})$ at the scale $\frac{\mu}{\lambda}$
\begin{align} \label{eq:expansion}
	G^{(n)}_{j_1 \ldots j_n }( x_1, \ldots, x_n; \mu, g(\frac{\mu}{\lambda}))\,\sim\, &{\mathcal G}^{(n,0)}_{j_1 \ldots j_n }( x_1, \ldots, x_n; \mu)+g^2(\frac{\mu}{\lambda}) \, {\mathcal G}^{(n,2)}_{j_1 \ldots j_n }( x_1, \ldots, x_n; \mu)\nonumber\\
	&+ g^4(\frac{\mu}{\lambda}) \, {\mathcal G}^{(n,4)}_{j_1 \ldots j_n }( x_1, \ldots, x_n; \mu)
	+\cdots\,.
\end{align}
Of course, the first term in the above expansion, being independent of the coupling, is the conformal contribution at zero coupling
\begin{equation}
	{\mathcal G}^{(n,0)}_{j_1 \ldots j_n }( x_1, \ldots, x_n; \mu) = G^{(n)}_{\text{conf} \, j_1 \ldots j_n }( x_1, \ldots, x_n) \, ,
\end{equation}
since the renormalized operators at zero coupling coincide with the conformal ones. 
The higher-order corrections in Eq. \eqref{eq:expansion} arise from the nonconformal interaction due to the nonvanishing beta function, so that the conformal contribution is corrected at higher orders in the renormalized coupling as displayed, the renormalized operators being nonconformal at higher orders in any renormalization scheme.
Yet, provided that the conformal contribution is nonvanishing, for fixed $x_1, \ldots, x_n$, all the higher-order terms in Eq. \eqref{eq:expansion} are suppressed with respect to the conformal one by powers of 
\begin{align} 
	g^2(\frac{\mu}{\lambda}) &\sim \dfrac{1}{\beta_0 \log(\frac{\mu^2}{\lambda^2\Lambda_{YM}^2})}\left(1-\frac{\beta_1}{\beta_0^2}\frac{\log\log(\frac{\mu^2}{\lambda^2\Lambda_{YM}^2})}{\log(\frac{\mu^2}{\lambda^2\Lambda_{YM}^2})}\right)\nonumber\\
	&\sim \dfrac{1}{\beta_0 \log(\frac{1}{\lambda^2})}\left(1-\frac{\beta_1}{\beta_0^2}\frac{\log\log(\frac{1}{\lambda^2})}{\log(\frac{1}{\lambda^2})}\right) \nonumber \\
\end{align}
-- i.e., asymptotically by inverse powers of $\log \frac{1}{\lambda}$ -- despite being in general nonconformal.
\\
Hence, the corresponding UV asymptotics, with fixed $x_1, \ldots, x_n$, reads \footnote{If $\mathcal{G}^{(n,0)}$ vanishes and $\mathcal{G}^{(n,2)}$ does not, the latter can be only computed on a case by case basis, since the YM action vertices would be involved in the computation. Therefore, finding an explicit  generalization of Eq. \eqref{eqrg} in this case is outside the scope of the present paper.} as $\lambda \rightarrow 0$
\begin{align} \label{eqrg}
	&\langle \mathcal{O}_{j_1}(\lambda x_1)\ldots\mathcal{O}_{j_n}(\lambda x_n)\rangle \,\sim \,\frac{Z_{\mathcal{O}_{j_1}}(\lambda)\ldots Z_{\mathcal{O}_{j_n}}(\lambda)}{\lambda^{D_{\mathcal{O}_1}+\cdots+D_{\mathcal{O}_n}}} G^{(n)}_{\text{conf}\,j_1 \ldots j_n }( x_1, \ldots, x_n)\,.
\end{align}
We refer to the aforementioned scheme as nonresonant diagonal \cite{Bochicchio:2021geometry}, whose existence involves the Poincar\'e-Dulac theorem via the following differential-geometric interpretation of operator mixing \cite{Bochicchio:2021geometry}.
We interpret a finite change of basis of the renormalized operators
\begin{equation}\label{linearcomb}
	\mathcal{O}'(x) = S(g) \mathcal{O}(x)
\end{equation}
in matrix notation as a real-analytic invertible gauge transformation $S(g)$ that depends on $g \equiv g(\mu)$. Then, the matrix
\begin{equation}
	A(g) = -\frac{\gamma(g)}{\beta(g)} = \frac{1}{g} \Big(\frac{\gamma_0}{\beta_0} + \cdots\Big)
\end{equation}
that enters the differential equation for $Z(\lambda)$
\begin{equation}
	\Big(\frac{\partial}{\partial g} - A(g)\Big) Z(\lambda) = 0
\end{equation}
defines a connection $A(g)$
\begin{eqnarray} \label{sys2}
	A(g)= \frac{1}{g} \left(A_0 + \sum^{\infty}_ {n=1} A_{2n} g^{2n} \right)\,,
\end{eqnarray}
with a regular singularity at $g = 0$ that transforms as
\begin{equation}
	A'(g) = S(g)A(g)S^{-1}(g) + \frac{\partial S(g)}{\partial g} S^{-1}(g)
\end{equation}
under the gauge transformation $S(g)$, with
\begin{equation}
	\mathcal{D} = \frac{\partial }{\partial g} - A(g)
\end{equation}
the corresponding covariant derivative.
Consequently, $Z(\lambda)$ is a Wilson line that transforms as
\begin{equation}
	Z'(\lambda) = S(g(\mu))Z(\lambda)S^{-1}(g(\frac{\mu}{\lambda}))\,.
\end{equation}
It follows from the Poincar\'e-Dulac theorem~\cite{Bochicchio:2021geometry} that, if any two eigenvalues $\lambda_1, \lambda_2, \ldots$ of the matrix $\frac{\gamma_0}{\beta_0}$, in nonincreasing order $\lambda_1 \geq \lambda_2 \geq \ldots$, do not differ by a positive even integer
\begin{equation} \label{nr}
	\lambda_i - \lambda_j - 2k \neq 0
\end{equation}
for $i \leq j$ and $k$ a positive integer -- i.e., they are nonresonant -- then a gauge transformation exists that sets $A(g)$ in the canonical nonresonant form \cite{Bochicchio:2021geometry}
\begin{equation} \label{1loop}
	A'(g) = \frac{\gamma_0}{\beta_0}\frac{1}{g}
\end{equation}
that is one-loop exact to all orders of perturbation theory. Hence, if in addition $\frac{\gamma_0}{\beta_0}$ is diagonalizable, Eq. \eqref{zz} follows.

\subsection{Nonresonant diagonal renormalization scheme  \label{app:C2}}

To make the present paper self-contained we provide the construction order by order in perturbation theory of the nonresonant diagonal scheme \cite{Bochicchio:2021geometry}.\par
The construction proceeds by induction on $k=1,2, \cdots$ by demonstrating that, once $A_0$ and the first $k-1$ matrix coefficients $A_2,\cdots,A_{2(k-1)}$ in Eq. \eqref{sys2} have been set in the canonical nonresonant form in Eq. \eqref{1loop} -- i.e., $A_0$ diagonal and $ A_2,\cdots,A_{2(k-1)}=0$ -- a real-analytic gauge transformation exists that leaves them invariant and sets the $k$-th coefficient $A_{2k}$ to $0$ as well. \par
The $0$ step of the induction consists in putting $A_0$ in diagonal form -- with the eigenvalues in nonincreasing order -- by a constant gauge transformation. \par
At the $k$-th step we choose the gauge transformation
\begin{eqnarray}
	S_k(g)=1+ g^{2k} H_{2k}\,,
\end{eqnarray}
with $H_{2k}$ a matrix to be found below. Its inverse reads
\begin{eqnarray}
	S^{-1}_k(g)= (1+ g^{2k} H_{2k})^{-1} = 1- g^{2k} H_{2k} + \cdots\,,
\end{eqnarray}
where the dots represent terms of order higher than $g^{2k}$.
The action of $S_k(g)$ on the connection $A(g)$ furnishes
\begin{align} \label{ind}
	&A'(g) \nonumber\\
	&=  2k g^{2k-1} H_{2k} ( 1- g^{2k} H_{2k})^{-1}+  (1+ g^{2k} H_{2k}) A(g)( 1- g^{2k} H_{2k})^{-1} \nonumber \\
	&=  2k g^{2k-1} H_{2k} ( 1- g^{2k} H_{2k})^{-1} +  (1+ g^{2k} H_{2k})  \frac{1}{g} \left(A_0 + \sum^{\infty}_ {n=1} A_{2n} g^{2n} \right)  ( 1- g^{2k} H_{2k})^{-1} \nonumber \\
	&=  2k g^{2k-1} H_{2k} ( 1- \cdots) +  (1+ g^{2k} H_{2k})  \frac{1}{g} \left(A_0 + \sum^{\infty}_ {n=1} A_{2n} g^{2n} \right)  ( 1- g^{2k} H_{2k}+\cdots) \nonumber \\
	&=  2k g^{2k-1} H_{2k}  +    \frac{1}{g} \left(A_0 + \sum^k_ {n=1} A_{2n} g^{2n} \right) + g^{2k-1} (H_{2k}A_0-A_0H_{2k}) + \cdots\nonumber \\
	&=   g^{2k-1} (2k H_{2k} + H_{2k} A_0 - A_0 H_{2k})  + A_{2(k-1)}(g) + g^{2k-1} A_{2k}+ \cdots\,,\nonumber \\
\end{align}
where we have skipped all the terms that contribute to an order higher than $g^{2k-1}$, with
\begin{align}
	A_{2(k-1)}(g) =  \frac{1}{g} \left(A_0 + \sum^{k-1}_ {n=1} A_{2n} g^{2n} \right)
\end{align}
that is the part of $A(g)$ that is not affected by the gauge transformation $S_k(g)$, and therefore verifies the hypotheses of the induction --  i.e., that $A_2, \cdots, A_{2(k-1)}$ vanish. \par
Thus, by Eq. \eqref{ind} the $k$-th matrix coefficient $A_{2k}$ may be eliminated from the expansion of $A'(g)$ to order $g^{2k-1}$ provided that an $H_{2k}$ exists such that
\begin{align}
	A_{2k}+(2k H_{2k} + H_{2k} A_0 - A_0 H_{2k}= A_{2k}+ (2k-ad_{A_0}) H_{2k}=0\,,
\end{align}
with $ad_{A_0}Y=[A_0,Y]$.
If the inverse of $ad_{A_0}-2k$ exists, the unique solution for $H_{2k}$ is
\begin{eqnarray}
	H_{2k}=(ad_{A_0}-2k)^{-1} A_{2k}\,.
\end{eqnarray}
Hence, to complete the induction, we should demonstrate that, if the eigenvalues of $A_0$ are nonresonant, $ad_{A_0}-2k$ is invertible, i.e., its kernel is trivial. \par
Now $ad_{\Lambda}-2k$, as a linear operator that acts on matrices, is diagonal, with eigenvalues $\lambda_{i}-\lambda_{j}-2k$ and the matrices $E_{ij}$, whose only nonvanishing entries are $(E_{ij})_{ij}$, as eigenvectors. The eigenvectors $E_{ij}$, normalized so that $(E_{ij})_{ij}=1$, form an orthonormal basis for the matrices. Therefore, $E_{ij}$ belongs to the kernel of $ad_{\Lambda}-2k$ if and only if $\lambda_{i}-\lambda_{j}-2k=0$. Consequently, since  $\lambda_{i}-\lambda_{j}-2k \neq 0$ for every $i,j$ by assumption, the kernel of $ad_{\Lambda}-2k$ only contains $0$, and the construction is complete.

\subsection{Anomalous dimensions of twist-$2$ operators  \label{app:C3}}

We define the bare collinear twist-$2$ operators with $s \geq 2$ and $k \geq 0$ \cite{Belitsky:1998gc}
\begin{equation}
	\mathcal{O}^{(k)}_{Bs} = (-i\partial_+)^{k}\mathcal{O}_{Bs}
\end{equation}
that, to the leading order of perturbation theory, for $k>0$ are conformal descendants \cite{Braun:2003rp} of the corresponding primary conformal operator 
$\mathcal{O}^{(0)}_{Bs}=\mathcal{O}_{Bs}$.
As a consequence of the operator mixing, we obtain \cite{Belitsky:1998gc,Braun:2003rp} for the renormalized operators
\begin{equation}\label{m}
	\mathcal{O}^{(k)}_{s}= \sum^{s}_{i} Z_{si} \mathcal{O}^{(k+s-i)}_{B i} \,,
\end{equation}
where $Z$ is the bare mixing matrix and 
\begin{equation}
	\gamma(g) =- \frac{\partial Z}{\partial \log \mu} Z^{-1}=\sum_{j=0}^{\infty}\gamma_j \, g^{2j+2}
\end{equation}
the -- lower triangular, in general -- matrix of the anomalous dimensions, with $\gamma_0$ diagonal in the $\overline{MS}$ scheme \cite{Belitsky:1998gc,Braun:2003rp}.
In our notation the eigenvalue of $\gamma_0$ corresponding to $\mathbb{O}_{s}$, with even $s\geq 2$, reads \cite{Belitsky:1998gc,Belitsky:2004sc}
\begin{align}
	\label{an1}
	\gamma_{0\mathbb{O}_{s}}= \frac{2}{(4 \pi)^2} \Bigg(&4 \psi(s+1) - 4 \psi(1) -\frac{11}{3}- 8 \frac{s^2+s+1}{(s-1)s(s+1)(s+2)}\Bigg) \,.
\end{align}
Consistently with the conservation of the stress-energy tensor $\gamma_{0\mathbb{O}_{2}}=0$, the remaining eigenvalues -- $\gamma_{0\mathbb{O}_{s}}$, with $s >2$ -- being positive. \par
Interestingly, from the above equations it follows that in SU($N$) YM theory $\gamma_0$ is independent of $N$ and thus has no nonplanar contribution \cite{Aglietti:2021bem}. \par 
We have verified numerically that the nonresonant condition in Eq. \eqref{nr} is satisfied up to $s=10^4$ by $	\gamma_{0\mathbb{O}_{s}}$ in Eq. \eqref{an1}. Moreover, subsequently the nonresonant condition for all the collinear twist-$2$ operators in YM theory has been proven \cite{S1}. \par 
Hence, the nonresonant diagonal basis exists for the collinear twist-$2$ operators in the present paper, and restricts to the lowest order of perturbation theory to the basis of conformal operators in Eq. \eqref{CO} -- where $\gamma_0$ is diagonal -- that may be employed to compute the conformal contribution in the right-hand side of Eq. \eqref{eqrg}.\par

\subsection{Generating functional of Euclidean RG-improved asymptotic correlators \label{app:C4}}

Finally, from the above construction of the Euclidean conformal generating functional and RG-improved correlators, it follows the generating functional of the Euclidean asymptotic correlators $\mathcal{W}^E_{\text{asym}}[J_{\mathbb{O}^E},\lambda]$ \cite{BPSpaper2}
\begin{align}
	\label{fullGenEexpZ}
	\mathcal{W}^E_{\text{asym}}[J_{\mathbb{O}^{E}},\lambda]
	=-(N^2-1)\log\Det\left(\mathbb{I}+\frac{1}{2}\frac{Z_{\mathbb{O}_{s}}(\lambda)}{\lambda^{s+2}}\Laplace^{-1}\frac{J_{\mathbb{O}^E_{s}}}{N}\otimes\mathcal{Y}_{s-2}^{E\,\frac{5}{2}}\right) \,,
\end{align}
according to Eqs. \eqref{WgenE1} and \eqref{eqrg}.
	
\newpage

\bibliographystyle{JHEP}
\bibliography{mybib} 

\providecommand{\href}[2]{#2}\begingroup\raggedright\begin{thebibliography}{10}

\bibitem{tHooft:1973alw}
G.~'t~Hooft, \emph{{A Planar Diagram Theory for Strong Interactions}},
  \href{http://dx.doi.org/10.1016/0550-3213(74)90154-0}{\emph{Nucl. Phys. B}
  {\bf 72} (1974) 461}.

\bibitem{Veneziano:1976wm}
G.~Veneziano, \emph{{Some Aspects of a Unified Approach to Gauge, Dual and
  Gribov Theories}},
  \href{http://dx.doi.org/10.1016/0550-3213(76)90412-0}{\emph{Nucl. Phys. B}
  {\bf 117} (1976) 519--545}.

\bibitem{Veneziano:1974dr}
G.~Veneziano, \emph{{An Introduction to Dual Models of Strong Interactions and
  Their Physical Motivations}},
  \href{http://dx.doi.org/10.1016/0370-1573(74)90027-1}{\emph{Phys. Rept.} {\bf
  9} (1974) 199--242}.

\bibitem{Aharony:1999ti}
O.~Aharony, S.~S. Gubser, J.~M. Maldacena, H.~Ooguri and Y.~Oz, \emph{{Large N
  field theories, string theory and gravity}},
  \href{http://dx.doi.org/10.1016/S0370-1573(99)00083-6}{\emph{Phys. Rept.}
  {\bf 323} (2000) 183--386}, [\href{https://arxiv.org/abs/hep-th/9905111}{{\tt
  hep-th/9905111}}].

\bibitem{Bochicchio:2017sgq}
M.~Bochicchio, \emph{{Renormalization in large-$N$ QCD is incompatible with
  open/closed string duality}},
  \href{http://dx.doi.org/10.1016/j.physletb.2018.06.072}{\emph{Phys. Lett. B}
  {\bf 783} (2018) 341--349},
  [\href{https://arxiv.org/abs/arXiv:1703.10176}{{\tt arXiv:1703.10176}}].

\bibitem{Gubser:1998bc}
S.~S. Gubser, I.~R. Klebanov and A.~M. Polyakov, \emph{{Gauge theory
  correlators from noncritical string theory}},
  \href{http://dx.doi.org/10.1016/S0370-2693(98)00377-3}{\emph{Phys. Lett. B}
  {\bf 428} (1998) 105--114}, [\href{https://arxiv.org/abs/hep-th/9802109}{{\tt
  hep-th/9802109}}].

\bibitem{Bochicchio:2017teh}
M.~Bochicchio, \emph{{The large-N Yang-Mills S-matrix is ultraviolet finite,
  but the large-N QCD S-matrix is only renormalizable}},
  \href{http://dx.doi.org/10.1103/PhysRevD.95.054010}{\emph{Phys. Rev. D} {\bf
  95} (2017) 054010}, [\href{https://arxiv.org/abs/arXiv:1701.07833}{{\tt
  arXiv:1701.07833}}].

\bibitem{Bochicchio:2016toi}
M.~Bochicchio, \emph{{An asymptotic solution of Large-N QCD, for the glueball
  and meson spectrum and the collinear S-matrix}},
  \href{http://dx.doi.org/10.1063/1.4949387}{\emph{AIP Conf. Proc.} {\bf 1735}
  (2016) 030004}.

\bibitem{streater2000pct}
R.~F. Streater and A.~S. Wightman, \emph{PCT, spin and statistics, and all
  that}, vol.~52.
\newblock Princeton University Press, 2000.

\bibitem{Migdal:1977nu}
A.~A. Migdal, \emph{{Multicolor QCD as Dual Resonance Theory}},
  \href{http://dx.doi.org/10.1016/0003-4916(77)90181-6}{\emph{Annals Phys.}
  {\bf 109} (1977) 365}.

\bibitem{Witten:1979kh}
E.~Witten, \emph{{Baryons in the 1/N Expansion}},
  \href{http://dx.doi.org/10.1016/0550-3213(79)90232-3}{\emph{Nucl. Phys. B}
  {\bf 160} (1979) 57--115}.

\bibitem{1987gauge}
A.~M. Polyakov, \emph{Gauge Fields and Strings}, vol.~3 of \emph{Contemporary
  concepts in physics}.
\newblock Taylor \& Francis, 1987.

\bibitem{Bochicchio:2013eda}
M.~Bochicchio, \emph{{Glueball and meson propagators of any spin in large-{N}
  {QCD}}}, \href{http://dx.doi.org/10.1016/j.nuclphysb.2013.07.023}{\emph{Nucl.
  Phys. B} {\bf 875} (2013) 621--649},
  [\href{https://arxiv.org/abs/arXiv:1305.0273}{{\tt arXiv:1305.0273}}].

\bibitem{Bochicchio:2023ols}
M.~Bochicchio, \emph{{Higher-Spin Currents, Operator Mixing and UV Asymptotics
  in Large-$N$ QCD-like Theories}},
  \href{http://dx.doi.org/10.3390/universe9020057}{\emph{Universe} {\bf 9}
  (2023) 57}.

\bibitem{BPSpaper2}
M.~Bochicchio, M.~Papinutto and F.~Scardino, \emph{{UV asymptotics of n-point
  correlators of twist-2 operators in SU(N) Yang-Mills theory}},
  \href{http://dx.doi.org/10.1103/PhysRevD.108.054023}{\emph{Phys. Rev. D} {\bf
  108} (2023) 054023}, [\href{https://arxiv.org/abs/arXiv:2208.14382}{{\tt
  arXiv:2208.14382}}].

\bibitem{BPS1}
M.~Bochicchio, M.~Papinutto and F.~Scardino, \emph{{n-point correlators of
  twist-2 operators in SU(N) Yang-Mills theory to the lowest perturbative
  order}}, \href{http://dx.doi.org/10.1007/JHEP08(2021)142}{\emph{JHEP} {\bf
  08} (2021) 142}, [\href{https://arxiv.org/abs/arXiv:2104.13163}{{\tt
  arXiv:2104.13163}}].

\bibitem{Bochicchio:2021geometry}
M.~Bochicchio, \emph{{On the geometry of operator mixing in massless QCD-like
  theories}},
  \href{http://dx.doi.org/10.1140/epjc/s10052-021-09543-5}{\emph{Eur. Phys. J.
  C} {\bf 81} (2021) 749}, [\href{https://arxiv.org/abs/arXiv:2103.15527}{{\tt
  arXiv:2103.15527}}].

\bibitem{harish1947infinite}
Harish-Chandra, \emph{Infinite irreducible representations of the {L}orentz
  group}, {\emph{Proceedings of the Royal Society of London. Series A.
  Mathematical and Physical Sciences} {\bf 189} (1947) 372--401}.

\bibitem{feldman1966unitarity}
G.~Feldman and P.~T. Matthews, \emph{Multimass fields, spin, and statistics},
  \href{http://dx.doi.org/10.1103/PhysRev.154.1241}{\emph{Phys. Rev.} {\bf 154}
  (1967) 1241--1249}.

\bibitem{streater1967local}
R.~F. Streater, \emph{Local fields with the wrong connection between spin and
  statistics}, \href{http://dx.doi.org/10.1007/BF01646839}{\emph{Communications
  in Mathematical Physics} {\bf 5} (1967) 88--96}.

\bibitem{Casalbuoni:2006fa}
R.~Casalbuoni, \emph{{Majorana and the Infinite Component Wave Equations}},
  \href{http://dx.doi.org/10.22323/1.037.0004}{\emph{PoS} {\bf EMC2006} (2006)
  004}, [\href{https://arxiv.org/abs/hep-th/0610252}{{\tt hep-th/0610252}}].

\bibitem{jaffe2006quantum}
A.~Jaffe and E.~Witten, \emph{Quantum {Y}ang-{M}ills theory}, {\emph{The
  millennium prize problems} {\bf 1} (2006) 129,
  http://www.claymath.org/prizeproblems/}.

\bibitem{wigner1939unitary}
E.~Wigner, \emph{On unitary representations of the inhomogeneous {L}orentz
  group}, {\emph{Annals of mathematics} (1939) 149--204}.

\bibitem{bochicchio2016glueball}
M.~Bochicchio, \emph{{Glueball and Meson Spectrum in Large-N QCD}},
  \href{http://dx.doi.org/10.1007/s00601-016-1100-6}{\emph{Few Body Syst.} {\bf
  57} (2016) 455--459}.

\bibitem{marino2015instantons}
M.~Marino, \emph{{Instantons and large N: an introduction to non-perturbative
  methods in quantum field theory}}.
\newblock Cambridge University Press, 2015.

\bibitem{maltoni2003color}
F.~Maltoni, K.~Paul, T.~Stelzer and S.~Willenbrock, \emph{{Color Flow
  Decomposition of QCD Amplitudes}},
  \href{http://dx.doi.org/10.1103/PhysRevD.67.014026}{\emph{Phys. Rev. D} {\bf
  67} (2003) 014026}, [\href{https://arxiv.org/abs/hep-ph/0209271}{{\tt
  hep-ph/0209271}}].

\bibitem{BPS3}
M.~Bochicchio, M.~Papinutto and F.~Scardino{\emph{{, to appear on arXiv}} }.

\bibitem{Belitsky:1998gc}
A.~V. Belitsky and D.~Mueller, \emph{{Broken conformal invariance and spectrum
  of anomalous dimensions in QCD}},
  \href{http://dx.doi.org/10.1016/S0550-3213(98)00677-4}{\emph{Nucl. Phys. B}
  {\bf 537} (1999) 397--442}, [\href{https://arxiv.org/abs/hep-ph/9804379}{{\tt
  hep-ph/9804379}}].

\bibitem{Belitsky:2004sc}
A.~V. Belitsky, S.~E. Derkachov, G.~P. Korchemsky and A.~N. Manashov,
  \emph{{Dilatation operator in (super-)Yang-Mills theories on the
  light-cone}},
  \href{http://dx.doi.org/10.1016/j.nuclphysb.2004.11.034}{\emph{Nucl. Phys. B}
  {\bf 708} (2005) 115--193}, [\href{https://arxiv.org/abs/hep-th/0409120}{{\tt
  hep-th/0409120}}].

\bibitem{Braun:2003rp}
V.~M. Braun, G.~P. Korchemsky and D.~Mueller, \emph{{The Uses of conformal
  symmetry in QCD}},
  \href{http://dx.doi.org/10.1016/S0146-6410(03)90004-4}{\emph{Prog. Part.
  Nucl. Phys.} {\bf 51} (2003) 311--398},
  [\href{https://arxiv.org/abs/hep-ph/0306057}{{\tt hep-ph/0306057}}].

\bibitem{Chan2021ModuliSO}
M.~Chan, \emph{Moduli spaces of curves: Classical and tropical},
  \href{http://dx.doi.org/https://doi.org/10.1090/noti2360}{\emph{Notices of
  the American Mathematical Society,} (2021) }.

\bibitem{Witten:2012ga}
E.~Witten, \emph{{Notes On Super Riemann Surfaces And Their Moduli}},
  \href{http://dx.doi.org/10.4310/PAMQ.2019.v15.n1.a2}{\emph{Pure Appl. Math.
  Quart.} {\bf 15} (2019) 57--211},
  [\href{https://arxiv.org/abs/arXiv:1209.2459}{{\tt arXiv:1209.2459}}].

\bibitem{pinch}
N.~Grulha~Jr., D.~Lima, K.~Rezende and M.~Zigart, \emph{{The effect of
  singularization on the Euler characteristic}},
  \href{http://dx.doi.org/10.21711/231766362023/rmc5312}{\emph{Matematica
  Contemporanea} {\bf 53} (2023) 254--277},
  [\href{https://arxiv.org/abs/2006.06056}{{\tt 2006.06056}}].

\bibitem{MBarxiv}
M.~Bochicchio{\emph{{, to appear on arXiv}} }.

\bibitem{Aglietti:2021bem}
U.~Aglietti, M.~Becchetti, M.~Bochicchio, M.~Papinutto and F.~Scardino,
  \emph{{Operator mixing, UV asymptotics of nonplanar/planar 2-point
  correlators, and nonperturbative large-N expansion of QCD-like theories}},
  \href{https://arxiv.org/abs/arXiv:2105.11262}{{\tt arXiv:2105.11262}}.

\bibitem{S1}
F.~Scardino, \emph{{Nonresonant renormalization scheme for twist-2 operators in
  SU(N) Yang{\textendash}Mills theory}},
  \href{http://dx.doi.org/10.1140/epjc/s10052-024-13590-z}{\emph{Eur. Phys. J.
  C} {\bf 84} (2024) 1229}, [\href{https://arxiv.org/abs/2410.15366}{{\tt
  2410.15366}}].

\end{thebibliography}\endgroup
\end{document}